\documentclass[a4paper,12pt]{article}
\usepackage[english]{babel}
\usepackage{amsmath,amssymb,amsbsy,amstext}
\usepackage[pdftex]{color,graphicx}

\usepackage{amsfonts}
\usepackage{amssymb}
\usepackage{mathrsfs}
\usepackage{epsfig}
\usepackage{subfigure}
\newcommand{\be}{\begin{eqnarray}}

\newcommand{\ee}{\end{eqnarray}}
\newcommand{\bdm}{\begin{displaymath}}
\newcommand{\edm}{\end{displaymath}}
   
\newcommand{\ba}{\begin{array}}
\newcommand{\ea}{\end{array}}

\newcommand{\sd}{\sigma_{up}}

\newcommand{\al}{\alpha'}

\newcommand{\zz}{z_{23}}
\newcommand{\xx}{x_{23}}
\newcommand{\esr}{\epsilon}
\newcommand{\bb}{b}

\newcommand{\g}{\tilde{g}}
\newcommand{\kq}{\kappa_4^2}
\newcommand{\Mp}{M_{Pl} }
\newcommand{\dd}{d}

\newcommand{\Vs}{V_{\Sigma}^w}

\begin{document}

\title{}
\author{}
\date{}
\thispagestyle{empty}

\begin{flushright}
\vspace{-3cm}
{\small LMU-ASC-09/08}\\
{\small arXiv:0802.2916}
\end{flushright}
\vspace{1cm}

\begin{center}
{\bf\LARGE
Inflation at the Tip}

\vspace{1.5cm}

{\bf Enrico Pajer\footnote{e.pajer@physik.uni-muenchen.de}}
\vspace{1cm}

{\it
Ludwig-Maximilians-University, Department of Physics \\ %
Theresienstr. 37, D-80333 M\"unchen, Germany 
}

\vspace{1cm}

{\bf Abstract}
\end{center}
\vspace{.5cm}
We study the motion of a (space filling) D3-brane at the tip of a warped deformed conifold, looking for inflationary trajectories. In our setup no anti D3-brane is present and the inflaton potential is induced by threshold corrections to the superpotential. First we study the slow roll regime and find that, allowing for fine tuning, hilltop inflation compatible with CMB data can take place. Then we consider the DBI regime and formulate a necessary condition for a phenomenologically viable inflationary stage. En passant, we propose a mechanism to cancel the large inflaton mass in the standard radial D3-anti D3-brane inflation.
\clearpage

%\listtoc \writetoc

\tableofcontents

%%%%%%%%%%%%%%%%%%%%%%%%%%%%%%%%%%%%%%%%%%%%%%%%%%%%%%%%%%%%%%%%%%%%%%%%%%%%%%%%

\section{Introduction}

The phenomenological success of the mechanism of inflation raises several basic questions such as for example which field plays the role of the inflaton, where does its potential come from and how does it couple to the standard model sector. This kind of questions can be addressed only in the context of a fundamental theory. In this note we investigate the possibility to embed inflation in string theory, in the framework of type IIB flux compactifications.

In particular, we study the motion of a (space filling) D3-brane on the tip of a warped deformed conifold looking for inflationary trajectories. Figure \ref{cartoon} gives a cartoon of the model. Before getting into the details of the construction and the results of the analysis, we would like to give some motivations to investigate such a model. 

Since its original proposal \cite{Dvali:1998pa}, brane inflation has been a popular setup for inflation in string theory. An important step forward was taken in \cite{Kachru:2003sx} (KKLMMT) where a D3-anti D3-brane system was embedded in a type IIB flux compactification \cite{Giddings:2001yu,Kachru:2003aw}. It was also realized there that a warped background can make the Coulomb potential suitable for a prolonged stage of inflation. Unfortunately, fixing all the (closed string) moduli induces, via the F-term, a mass for the inflaton that, if not cancelled by some other effects, prevents slow roll inflation.

In the KKLMMT model, the anti D3-brane plays a threefold role: it generates the inflaton potential, it annihilates with the inflating D3-brane providing a reheating mechanism\footnote{Actually the tachyon condensation driving the annihilation process might be as well responsible for graceful exit from inflation. This is very model dependent. E.g. in the setup analyzed in \cite{Lorenz:2008pc}, it was found that the CMB data prefer the end of inflation due to the failure of the slow roll conditions than due to tachyon condensation.} and finally it breaks SUSY and can provide an effective four dimensional de Sitter (or Minkowski) space. Let us separately analyze these issues and emphasize how the \textit{inflation at the tip} model we propose in this letter addresses them in a different way.
\begin{enumerate}
\item Because of the gravitational backreaction of the mobile D3-brane, once the volume (K\"ahler) modulus has been stabilized, a non trivial potential for the D3-brane position is generically produced. If an anti D3-brane is present, this \textit{stabilization potential} competes with the Coulomb potential to determine the inflaton dynamics. Several attempts have been made to find a model where, due to some symmetry or fine tuning of the parameters, the stabilization potential gives a negligible contribution. The motivation of these attempts was that, contrary to the relatively simple Coulomb potential, the stabilization effects, encoded in the F-term, depend generically in a complicated way on the details of the compactification\footnote{Another interesting proposal \cite{Alishahiha:2004eh} comes from noticing that the non standard (DBI) D-brane kinetic term allows a prolonged stage of inflation even for steep potentials. In the following we argue that also from this point of view, i.e. considering DBI inflation, inflation at the tip gives a particularly simple model.}. In the present paper, we would like to take the opposite point of view, i.e. we propose a simple model where the inflaton potential is determined exclusively by moduli stabilization effects. One clear advantage is that it is easy to imagine a model where anti D3-branes are absent or sufficiently far away. Then no additional symmetries or fine tuning are required at least a priori. 
\begin{figure}
\centering
\includegraphics[width=0.5\textwidth]{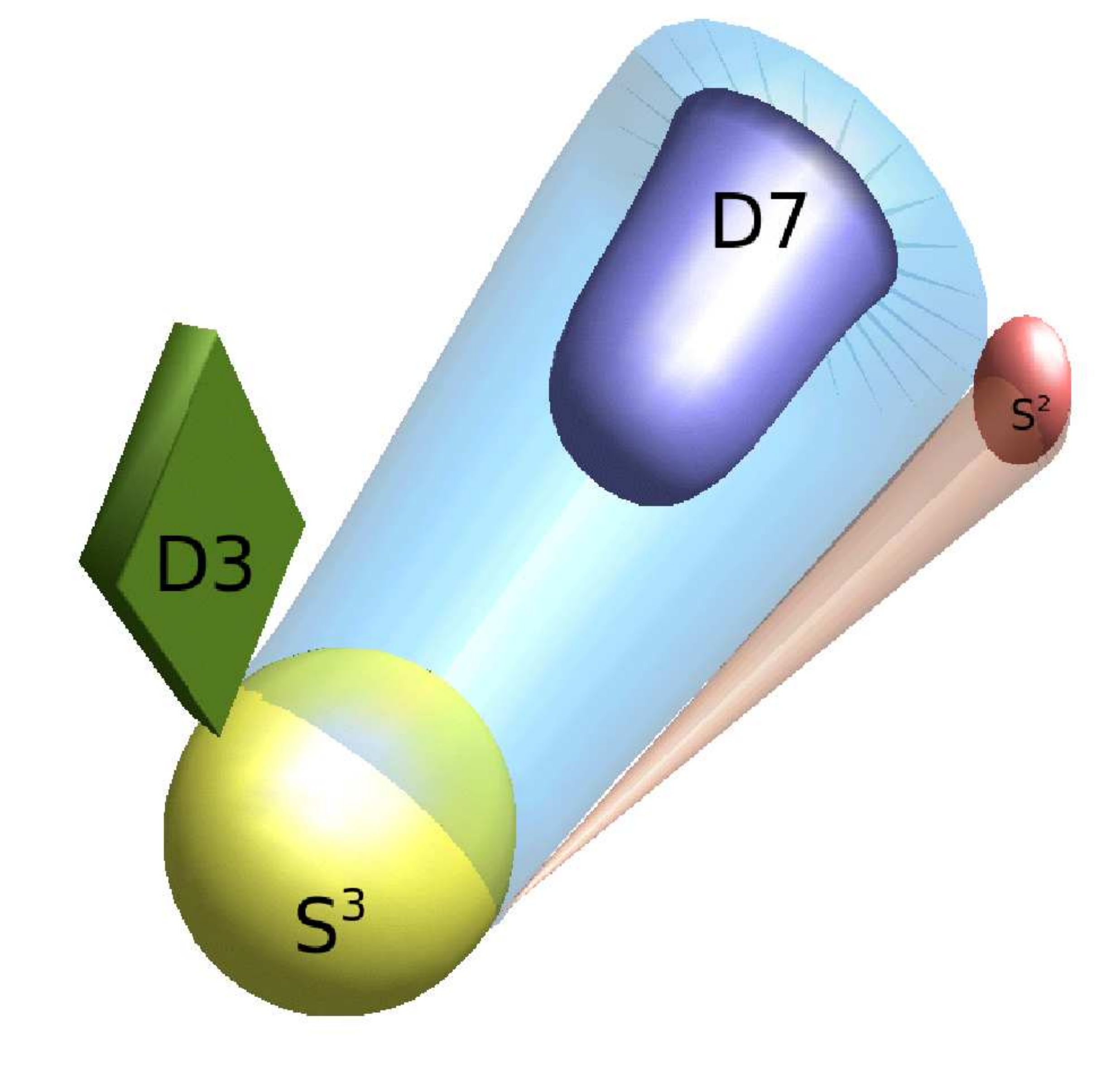} \hspace{-1cm}
%\includegraphics[width=0.5\textwidth,height=0.4\textwidth]{cartoon2}
%\caption{A cartoon of inflation at the tip: on the left the simplest model with just one D3-brane and a very symmetric D7 embedding; this is the setup that we thoroughly study in this paper. On the right a possible generalization is depicted, with many D3-branes and a more complicated D7 embedding.}
\caption{A cartoon of the inflation at the tip model where we zoom in close to the tip of the warped throat.\label{cartoon}}
\end{figure}
\item Without an anti D3-brane, a different mechanism from brane annihilation has to be responsible for reheating. Although we do not address this issue here, we would like to make a few remarks. One of the original motivations for brane inflation \cite{Dvali:1998pa} was to obtain an efficient reheating mechanism; the interesting feature of brane brane collision (contrary to annihilation) is that after the collision the kinetic energy of the relative motion is almost entirely transferred into modes living on the newly created stack of branes and the creation of bulk (Kaluza Klein) modes is negligible. The process of brane collision has been carefully studied \cite{Kofman:2004yc} and this expectation have been confirmed. On the other hand, a thorough phenomenological analysis of this reheating mechanism has not yet been performed. It would be very interesting to have a quantitative estimate of the efficiency of this reheating process. For inflation at the tip with many D3-branes (a straightforward generalization of the single D3-brane case), the end of inflation is precisely the collision of several of them (in the common minimum of the potential); this reheating mechanism (to which we will refer to as ``brane trapping'') would hence be naturally embedded.
\item In addition to the original anti D3-brane at the tip of a warped throat \cite{Kachru:2003aw}, several other mechanisms have been proposed to break SUSY and uplift the effective four dimensional AdS vacuum to a dS one. We will therefore try to keep our analysis as general as possible, without specifying the uplifting mechanism; we will find at the end that some constraints on the scaling of the uplifting are indeed present.
\end{enumerate}
% \begin{figure}
% \centering
% %\includegraphics[width=0.4\textwidth,height=0.2\textwidth]{coulomb.eps}
% \caption{The Coulomb plus gravitational potential between a D3 and an anti D3-brane becomes extremely steep for small brane separation.\label{fig:coulomb}}
% \end{figure}

Another motivation to consider inflation at the tip has to do with the direction along which the D3-brane moves. In the standard setup \cite{Kachru:2003sx}, the motion takes place along the radial direction of a warped conifold. The radial position comes out to be a conformally coupled scalar in the four dimensional effective action, so that it is typically a direction too steep for slow roll inflation. This motivates the study of angular motion\footnote{Works who studied inflation driven by an exclusively angular motion of branes are \cite{DeWolfe:2004qx,DeWolfe:2007hd}, who found a negative result. In \cite{Easson:2007dh}, the cosmological implications of a combined angular and radial motion were considered.}. In \cite{DeWolfe:2007hd} for example, the possibility of slow roll inflation for D3-anti D3-branes at the tip, i.e. separated only in the angular directions, was considered. The authors found a negative result that a posteriori can be understood as follows. The brane anti brane separation is bounded from above by the size of the tip that is extremely small. The Coulomb potential becomes very steep for small brane separation which makes slow roll inflation impossible. For the model we propose in this paper the potential depends on the compactification (and it is not ``universal'' as the Coulomb potential) and can therefore be made flat with an appropriate choice of the parameters (if and when this requires fine tuning will be discussed further in the paper). 

A final motivation is related to DBI inflation. The idea of \cite{Alishahiha:2004eh} is that, although the inflaton potential is not flat (in the slow roll sense), inflation can nevertheless last long enough because the DBI kinetic term imposes an upper bound on the inflaton speed. The DBI kinetic term is determined by the warp factor that, for the warped conifold, is a function of the radial position. During inflation at the tip (constant radius) the warp factor is constant, which gives, in a certain sense, the simplest DBI kinetic term possible. This makes the inflationary analysis particularly easy.

The model we study (schematically depicted in figure \ref{cartoon}) consists of a space filling D3-brane moving along the tip of a warped deformed conifold which is an $S^3$. The potential depends on three scalar fields (the three angles of $S^3$) and comes from the F-term which is determined by the supersymmetric embeddig of a stack of (space filling) D7-branes. The choice of a particularly symmetric D7 embedding gives rise to a very simple scalar potential. In different regions of the parameter space this potential can give rise to different cosmological evolutions. For example, we will show that a quite general choice of parameters leads to the natural inflation potential; in some fine tuned regions one can also obtain some kind of hilltop potential. Given the simplicity of the potentials we find, we do not need to rely on numerical methods for the inflationary analysis and also for the DBI inflation important results can be obtained analytically.

First we look for slow roll inflation and we find that, fine tuning stringy parameters, the potential can be made flat enough. Phenomenologically, this corresponds to a model is of the hilltop type \cite{Kohri:2007gq}, with $n_s\simeq0.94$, negligible tensor modes and an adjustable scale of inflation $\Lambda^{1/4}\sim d\times 10^{-3}$ where $d$ is given in \eqref{can} and is generically subplanckian. Then we look for DBI inflation and we find that it is not possible in the simplest model. In fact, we show that DBI inflation can produce the right perturbations (with interesting non Gaussianity signature) but not enough e-foldings. Then we argue that, considering more generic embeddings (as depicted in figure \ref{recy}), one can obtain a viable inflationary model in which DBI and slow roll phases alternate. For example the perturbations can be produced in the DBI regime, while the rest of the $60$ observable e-foldings take place in the slow roll regime. We leave the detailed study of these DBI-slow roll alternating models for future work.

There are two other results that came out of the present investigation, that are not aligned with the main focus of this paper. One regards Natural Inflation. As we discuss in section \ref{subsec:nut1}, the phenomenological constraint that the axion decay constant has to take superplanckian values \cite{Savage:2006tr} is relaxed in presence of a non canonical kinetic term. We study explicitly the case of a DBI kinetic term. The second result consists in noticing that the uplifting term plays a crucial role for the radial dynamics of a brane in a warped conifold. As we explain in appendix \ref{app:up}, this can be used to cancel the mass of the inflaton in the standard radial D3-anti D3-brane inflation. 

The paper is organized as follows. In section \ref{ST} we set the stage and calculate the scalar potential for a generic embedding. In section \ref{sec:kup} we specify the D7 embedding and find an explicit potential; we discuss the stabilization of the other moduli (K\"ahler modulus and radial displacement) and introduce the uplifting. The final result of this first part is the potential appearing in \eqref{pot no}. This is the starting point for the inflationary analysis of section \ref{sec:inlf}; we consider both the slow roll and the DBI regimes. In section \ref{no go} we generalize our analysis of DBI inflation. We present a no go result and describe how to evade it. We finish with some comments, conclusions and perspectives in section \ref{sec:con}.

%%%%%%%%%%%%%%%%%%%%%%%%%%%%%%%%%%%%%%%%%%%%%%%%%%%%%%%%%%%%%%%%%%%%%%%%%%%%%%%%%%%%%%%%%%%%%%%%%%%%%%%%%%%

\section{F-term potential for a D3-brane at the tip}\label{ST}

In this and the next section we calculate the potential for a single D3-brane moving at the tip of a warped deformed conifold \cite{Klebanov:2000hb} in the framework of type IIB flux compactification. In this section we set the stage and obtain for the potential the general result \eqref{pot}. In section \ref{sec:kup} we consider a simple embedding and obtain the more explicit result \eqref{pot no}.

Following \cite{Kachru:2003aw}, we assume that the complex structure moduli and the dilaton have been stabilized by fluxes \cite{Giddings:2001yu}. We make the simplifying assumption that there is only one K\"ahler modulus $\rho$. The $N=1$ supergravity scalar potential is given by
\be
    V_F = e^{\kq K}\left ( K^{\bar I J}D_J W\overline{D_I W}
    -3\kappa_4^2|W|^2\right)
\ee
where the indices $I,J$ run over the K\"ahler modulus $\rho$ and the open string moduli $z_i$ with  $i=1,2,3$ describing the D3-brane position. The K\"ahler potential $K$ is given by \cite{DeWolfe:2002nn}
\be\label{K}
\kq K=-3\mathrm{log}[\rho+\bar \rho-\gamma k(z,\bar z) ]\equiv-3\mathrm{log}U,
\ee
where $\gamma$ is a constant and $k$ is the K\"ahler potential of the Calabi Yau evaluated at the position $z$ of the D3-brane. If a warped throat \cite{Klebanov:2000hb} is present, deep inside it, $k$ is well approximated by the K\"ahler potential of the deformed conifold.

The conifold is defined by the following hypersurface in $\mathbb{C}^4$
\be\label{coni}
\sum_{A=1}^4(z^A)^2=\varepsilon^2\,,
\ee 
and its tip by
\be
\sum_{A=1}^4|z^A|^2=\varepsilon^2\,.
\ee
Putting these equations together and writing them in terms of $z_A=x_A+i y_A$, with $A=0,1,2,3$, one finds that the tip is an $S^3$ embedded in a real slicing of the $z_A$ complex coordinates
\be\label{tip}
\mathrm{tip:}\qquad\sum_{A=1}^4x_A^2=\varepsilon^2\,,\qquad y_A=0\,.
\ee
Close to the tip, the K\"ahler potential of the deformed conifold takes the form \cite{Candelas:1989js,DeWolfe:2007hd}
\be\label{kahler}
k(z,\bar z)=k_0+c\varepsilon^{-2/3}\left( \sum_{A=1}^{4}|z_A|^2-\varepsilon^2 \right)
\ee
where $c=\frac {2^{1/6}}   {3^{1/3} }\simeq 0.77$. Hence, at the tip $k(z,\bar z)$ is just a constant; this implies that $U=\rho+\bar\rho-\gamma k_0$ does not depend on the position of the brane on the tip. Also, one can verify \cite{DeWolfe:2007hd} that at the tip $k$ is stationary in all directions: $\partial_{z^A} k=0$. Thanks to this property, the K\"ahler metric and its inverse \cite{Burgess:2006cb} take a simple block diagonal form:
\be\label{Kij}
    K_{I \bar J} &=& \frac{3}{\kappa^2U^2}
    \left(\begin{array}{cc} 1 & -\gamma k_{\bar j} \\
    -\gamma k_{i} & U \gamma k_{i \bar j}+\gamma^2 k_{i} k_{\bar j}
    \end{array}\right)=
\frac{3}{U^2\kappa^2}
    \left(\begin{array}{cc} 1 & 0 \\
    0 & U \gamma k_{i \bar j}
    \end{array}\right)\\
\label{Kji} K^{\bar I J} &=& \frac{\kappa^2U}{3}
    \left(\begin{array}{cc} U+\gamma k_{\bar l}k^{\bar{l}h}k_h &k_{\bar l} k^{\bar{l}j}  \\
   k^{\bar{i}l} k_{l}  & \gamma^{-1} k^{ \bar i j}
    \end{array}\right)=
\frac{\kappa^2 U}{3}
    \left(\begin{array}{cc} U & 0 \\
    0 & \gamma^{-1} k^{\bar i j}
    \end{array}\right)\,.
\ee
Choosing ${z_1,z_2,z_3}$ as independent coordinates and $z_4^2=\varepsilon^2-\sum_{i=1}^3z_i^2$, from \eqref{kahler} we obtain
\begin{equation}
\begin{array}{rclcrcl}
k_{i\bar j}&=&\frac{ c}{\varepsilon^{2/3}}\left(\delta_{i\bar j} +\frac{z_i \bar z_j}{|z_4|^2}\right) &\longrightarrow& k_{i\bar j}&=&\frac{c}{\varepsilon^{2/3}}\left(\delta_{i\bar j} +\frac{x_i x_j}{\varepsilon^2-\sum x_h^2}\right)  \nonumber\\
&&&\mathrm{at\, the\, tip}& \nonumber\\
k^{\bar i j}&=&\frac{\varepsilon^{2/3}}{c} \left(\delta^{\bar i j} -\frac{z_i \bar z_j}{\sum |z_A|^2}\right) &\longrightarrow& k^{\bar i j}&=&\frac{\varepsilon^{2/3}}{c} \left(\delta^{\bar i j} -\frac{x_i x_j}{\varepsilon^2}\right) \label{inv} \\
\end{array}
\end{equation}
where on the right side we used that $z_A=\bar{z}_A=x_A$.

A stack of $n$ D7-branes wraps a divisor $\Sigma$ defined by the zero of a holomorphic function\footnote{We use $g$ instead of $f$ as in \cite{Burgess:2006cb,Baumann:2006th,DeWolfe:2007hd,Krause:2007jk,Baumann:2007ah} to avoid confusion with the DBI parameter in \eqref{action}.} $g(z)$. Gaugino condensation can generate a non perturbative term $W_{np}$ in the $N=1$ superpotential in addition to the Gukov Vafa Wittten term $W_0$ \cite{Gukov:1999ya}:
\be \label{sup} 
W&=&W_0+W_{np}=W_0+A\,e^{-a\rho }\,.
\ee
The prefactor $A$ appearing in $W_{np}$ depends in principle on open string, the complex structure moduli and the dilaton. As we assumed the latters to be fixed by fluxes, we concentrate on the former. In \cite{Ganor:1998ai}, Ganor argued that $A$ has to depend on the open string moduli, that we indicate generically with $z$ here, in such a way that it vanishes when the D3-brane is on $\Sigma$. In \cite{Berg:2004ek} $A$ has been computed in the case of toroidal orientifolds. In \cite{DeWolfe:2007hd,Baumann:2006th} it was shown how to generalize the calculation to curved space using the Green function method. For the singular conifold and $Y^{(p,q)}$ cones the result \cite{Baumann:2006th} is 
\be \label{conj}
A=A_0 \,g(z)^{1/n}
\ee
where $A_0$ depends just on the complex structure moduli and the dilaton and will be treated as a constant in the following. The consistency with Ganor's argument led to the conjecture that \eqref{conj} is valid for a generic compactification \cite{Baumann:2006th}.

From a different perspective, another argument in favor of \eqref{conj} was formulated in \cite{Koerber:2007xk} using the results of \cite{Martucci:2006ij}. The idea of the argument is that, from the 10 dimensional point of view, the presence of non perturbative corrections to the four dimensional superpotential means that the 10 dimensional Einstein equations do not require any more an ordinary complex structure but a generalized complex structure \cite{Koerber:2007xk}. The latter determines then a non trivial superpotential for the D3-brane (that is vanishing in the warped Calabi Yau case). Using the potential of \cite{Martucci:2006ij} and Ganor's argument \cite{Ganor:1998ai}, one can obtain \eqref{conj}.

Therefore, for the deformed conifold we will use the superpotential
\be
W=W_0+A_0\, g(z)^{1/n} \, e^{-a\rho}\,.
\ee
The F-term potential for $K$ given in \eqref{K}, using \eqref{Kij} and \eqref{Kji}, takes the form
\be
    V_F &=& \frac{\kappa_4^2}{ 3 U^2}\left[ U|W_{,\rho}|^2 -3(\bar{W} W_{,\rho} + {\rm c.c.} )
   +\frac{1}{\gamma} k^{\bar i j} \overline W_{,\bar i} W_{,j}\right] \,.
\ee
For the superpotential in \eqref{sup}, this becomes
\be
    V_F= \frac{\kappa_4^2}{ 3 U^2}\Bigg[ \left(Ua^2+6a\right)
    |A|^2e^{-a(\rho+\bar\rho)} + 3a(\bar{W_0} Ae^{-a\rho} +\mathrm{c.c.})  \\
\hspace{4cm}  
  + \frac{1}{\gamma}  k^{\bar i j} \overline A_{,\bar i} A_{,j}
    e^{-a(\rho+\bar\rho)}\Bigg] \,.  \nonumber
\ee
The minimization of the axion works as usual (see e.g. \cite{Krause:2007jk}), leading to the final result
\be \label{pot}
V_F&=& V_{KKLT}+\Delta V \nonumber\\
&=&\frac{\kappa_4^2 2 a |A| e^{-a\sigma} } { U^2} \left( \frac16 a U |A| e^{-a\sigma} + |A| e^{-a\sigma} -|W_0| \right)\\
\nonumber &&+\frac{\kappa_4^2 e^{-2a\sigma}}{3U^2 \gamma}  k^{\bar i j} \overline A_{,\bar i} A_{,j}\,.
\ee
We would like to stress the fact that this form of the potential is valid only for the D3-brane at the tip, where the K\"ahler metric \eqref{Kji} is block diagonal. In this case the potential depends on the variables $\sigma,x_1,x_2,x_3$.
In \cite{DeWolfe:2007hd}, the supersymmetric vacua of this potential were studied solving \mbox{$D_IW=0$} for two classes of embeddings. Depending on the choice of the embedding, the set of supersymmetric vacua can be empty, a point, one or two dimensional. 

The symmetries of the problem give us an important insight. With $W_{np}$ set to zero, the F-term potential for a D3-brane moving along the tip (where $k=k_0$) enjoys an $SO(4)$ symmetry acting on the coordinates $z_A=x_A$, i.e. the symmetry of $S^3$. The nonperturbative effects break this symmetry via the embedding function $g(z)$. An appropriate choice of $g(z)$ can lead to a very symmetric and simple potential $V$. In the next section we provide an explicit example of this where $g(z)$ breaks $SO(4)$ to $SO(3)$ and the potential $V$ depends on the D3-brane position $z_A=x_A$ only via a single real field $\phi$. In section \eqref{no go} we comment on the implications of a generic $g(z)$.

%%%%%%%%%%%%%%%%%%%%%%%%%%%%%%%%%%%%%%%%%%%%%%%%%%%%%%%%%%%%%%%%%%%%%%%%%%%%%%%%%%%%%%%%%%%%%%%%%%%%%%%%%%%%%%%

\section{Kuperstein embeddings}\label{sec:kup}

We now calculate the potential in \eqref{pot} for a class of particularly symmetric embeddings. We also review the result of the radial stabilization (performed in appendix \ref{app:radial}) that guaranties that the D3-brane at the tip is at a local minimum in the radial direction. Our final result is \eqref{pot no} (valid for large volume and when $\varepsilon/\mu\gg1$, i.e. when the D7's are not too close to the tip), which will be the starting point for the inflationary analysis of the next section. 

As we already said, the tip of the KS throat is an $S^3$ which can be described as the real slicing of the $z_A$ coordinates plus the constraint \eqref{tip}. The $SO(4)$ symmetry of $S^3$, acting naturally on ${x_1,x_2,x_3,x_4}$, is broken once a certain embedding function $g$ (defining the divisor $\Sigma$ where the stack of D7-brane is wrapped) is chosen. 

A family of supersymmetric embeddings was found by Kuperstein in \cite{Kuperstein:2004hy}. It is given, up to $SO(4)$ permutations, by
\be \label{kup}
g(z)=\g \left( z_2^2+z_3^2 \right)-z_1=\g(\zz)-z_1 \,,
\ee
where $\g(\zz)$ is a holomorphic function of its argument $\zz\equiv z_2^2+z_3^2$. For a generic $\g(x)$ these embeddings preserve an $SO(2)$ symmetry corresponding to rotations in the $z_2-z_3$ plane. The first term in the potential, $V_{KKLT}$ is easily calculated substituting $A=A_0 g(x)^{1/n}$ in \eqref{pot}. For the term $\Delta V$, a straightforward calculation using \eqref{inv} leads to
\be \label{gen}
\Delta V&=& \frac{\kq |A(x)|^2e^{-2a\sigma}\varepsilon^{2/3}      }{3cn^2U^2\gamma |\g-x_1|^2}\cdot \\
\nonumber &&\qquad \cdot \left[ 1-\frac{x_1^2}{\varepsilon^2}+4\xx \, \g'^2\left(1-\frac{\xx}{\varepsilon^2}\right)+4\g'\frac{\xx x_1}{\varepsilon^2}  \right]\,,
\ee
where the prime indicates derivative with respect to $\zz$.

%%%%%%%%%%%%%%%%%%%%%%%%%%%%%%%%%%%%%%%%%%%%%%%%%%%%%%%%%%%%%%%%%%%%%%%%%%%%%%%%%%%%%%%%%%%%%%%%%%%%%%%%%%%%%%%

\subsection{A simple case}\label{sec:simpl}

For certain choices of $\g$, the potential can exhibit very flat regions (e.g. inflection points); the difficulty is that multi inflaton analysis is in general required. In section \eqref{no go} we will further comment on a generic $\g$, but a thorough analysis is left for future investigation. 

\begin{figure}
\centering
\includegraphics[width=0.8\textwidth]{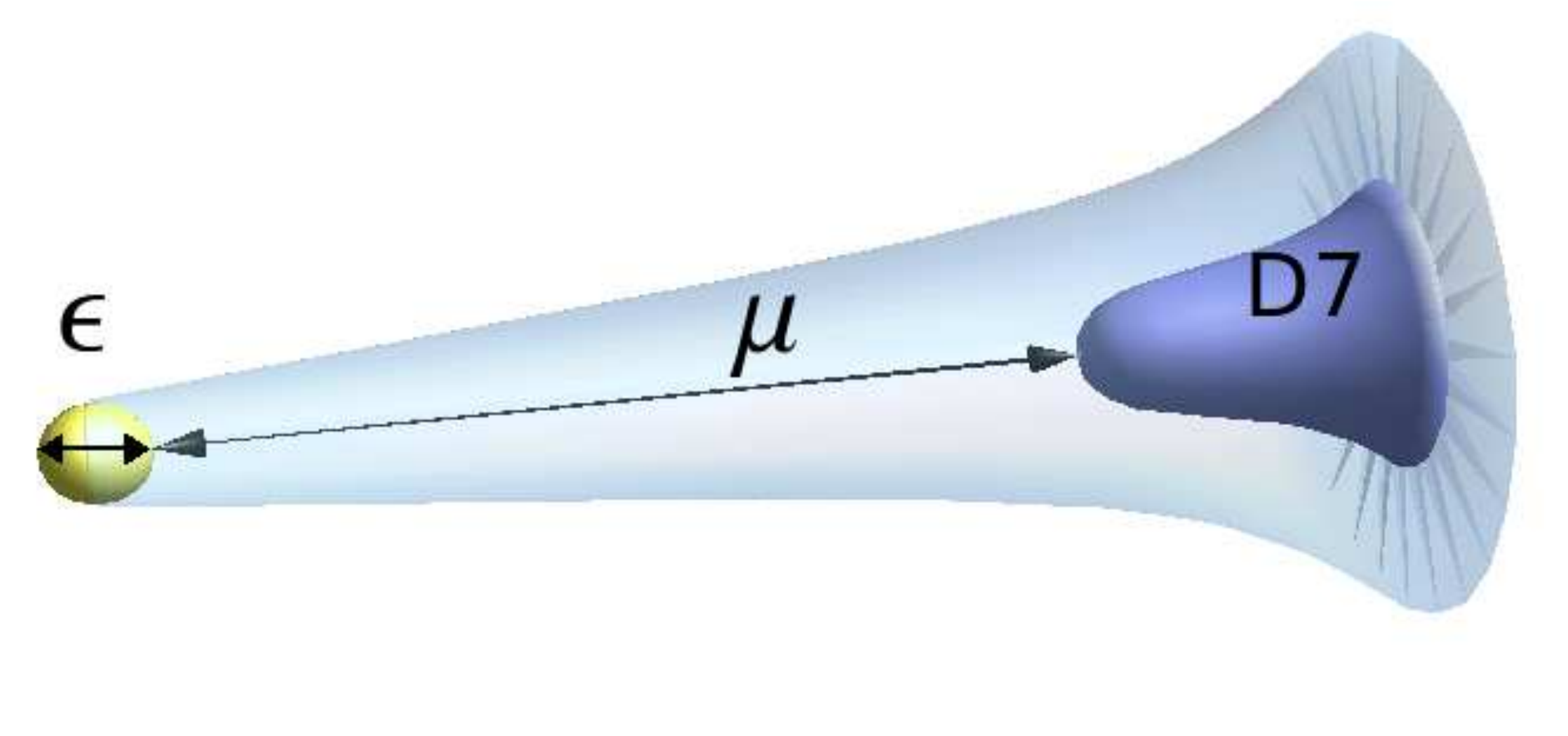}
\caption{The drawing describes the geometrical meaning of the parameters $\mu$, i.e. the distance of the stack of D7-branes to the tip, and $\varepsilon$, i.e. the size of the tip.\label{fig:mu}}
\end{figure}

In this section, we focus on the simplest case when $\g(\zz)=\mu$, where without lost of generality we can take $\mu$ real and positive. As depicted in figure \ref{fig:mu}, this parameter indicates the deepest radius reached by the stack of D7-branes $r_{D7}=\mu^{2/3}$. We are interested in the case where the D7's do not reach close to the tip, such that $\varepsilon/\mu\ll1$ is a good expansion parameter. The non perturbative superpotential is 
\be\label{np kup}
W_{np}=A(z)e^{-a\rho}=A_0\left(1-\frac{z_1}{\mu} \right)^{1/n}e^{-a\rho}\,,
\ee
where we have rescaled $A_0$ by a factor $\mu^{1/n}$, so that $A(z)=A_0+\mathcal{O}(\varepsilon/\mu)$. As now $g(z)=g(z_1)$, the $SO(4)$ symmetry of the potential is broken to $SO(3)$ that acts naturally on ${z_2,z_3,z_4}$.

Using \eqref{np kup}, the potential \eqref{pot} becomes
\be \label{pot2}
V_{KKLT}&=&\frac{\kappa_4^2 2 a e^{-a\sigma} } { U^2} \left[ \frac16 a U |A(x_1)|^2 e^{-a\sigma} + |A(x_1)|^2 e^{-a\sigma} -|W_0A(x_1)| \right]\\
\Delta V&=&\frac{\kq e^{-2a\sigma}}{3U^2 \gamma}  k^{\bar 1 1} \overline A(x_1)_{,\bar 1} A(x_1)_{,1} \label{pot3}\\
&=&\frac{\kq |A(x_1)|^2 e^{-2a\sigma}}{3cn^2\mu^2 U^2 \gamma}  \varepsilon^{2/3}\left(1 -\frac{x_1^2}{\varepsilon^2}\right) \left(1-\frac{x_1}{\mu}\right)^{-2} \,,\nonumber
\ee
and depends only on two real variables $\sigma$ and $x_1$. Thanks to the $SO(3)$ symmetry, two of the three independent coordinates $x_i$ with $i={1,2,3}$ are flat directions of the scalar potential, i.e. $x_2$ and $x_3$. Some effects, such as for example corrections from the bulk\footnote{Notice that, as argued in \cite{DeWolfe:2007hd}, bulk effects are in general subleading with respect to the potential we are considering here. They are important only when the potential has a completely flat direction, as it is the case for the $x_2$ and $x_3$ directions.}, will eventually stabilize these directions. In the inflationary analysis of section \ref{sec:inlf} we will therefore assume that the D3-brane starts and stays at a minimum in the $x_2$ and $x_3$ directions and moves only along $x_1$. This technical assumption is almost ubiquitous in the brane inflation literature. The effects of these ``flat'' directions can be investigated in a second step once a successful model of inflation is found.

In principle, the kinetic term $K_{i\bar j}\partial z_i \partial \bar z_{j}$ mixes all the $z_i$'s. For our case, a more convenient parameterization of the $S^3$ is
\begin{displaymath}
\begin{array}{rclrcl}
z_1&=x_1=&\varepsilon\cos\phi  \,,  &   z_2&=x_2=&\varepsilon \cos\phi \,\sin \psi\,\sin\theta\,, \\
z_3&=x_3=&\varepsilon \sin\phi \,\sin\psi\,\sin\theta   \qquad , &  z_4&=x_4=&\varepsilon\cos\theta\,\sin\psi \,, \\
\end{array}
\end{displaymath}
where $\phi$ runs from $0$ to $\pi$. This choice leads to the diagonal metric
\be 
ds^2=c\varepsilon^{4/3}\left[d\phi^2+\sin^2 \phi (d\theta^2+\sin^2\theta d\psi^2)\right]\,.
\ee
In these new coordinates, the potential in \eqref{pot2} and \eqref{pot3} is just a function of $\sigma$ and $\phi$ and the kinetic term is diagonal. 

We write now the potential in these new coordinates. We expand $V_{KKLT}$ and $\Delta V$ for $\varepsilon/\mu\ll 1 $:
\be \label{ex}
V_{KKLT}&\simeq&\frac{2\kq |A_0|ae^{-a\sigma}}{U^2}\left(\frac16 |A_0|aUe^{-a\sigma}+|A_0|e^{-a\sigma}-|W_0|\right)\nonumber\\
&&+\frac{2\kq \varepsilon |A_0|ae^{-a\sigma}}{U^2n\mu}\left(\frac13 |A_0|aUe^{-a\sigma}+2|A_0|e^{-a\sigma}-|W_0|\right)\,\cos\phi+\dots \nonumber\\
\Delta V&\simeq& \frac{\kq|A_0|^2 e^{-2a\sigma} \varepsilon^{2/3}}{3c n^2 \mu^2 U^2 \gamma } \sin^2\phi+\dots
\ee

We know that the first term in \eqref{ex}, gives rise to an AdS minimum \cite{Kachru:2003aw} (that can be trusted for small $W_0$) and that the other terms are suppressed with respect to the first by a factor $\varepsilon /(n\mu)$ and $\varepsilon^{2/3}/(n\mu)^2$ respectively. Hence if we want to have inflation we need an uplifting. There are several possibilities in the literature to obtain it: an anti D3-brane as in \cite{Kachru:2003aw} (but in this case it will have to be located far away, in another throat, so that the attractive force can be neglected), a D-term \cite{Burgess:2003ic} or an F-term uplifting \cite{Brax:2007xq,Saltman:2004sn}. To keep our analysis general, in the following we will assume that a term 
\be \label{Vup}
V_{up}=\frac{D}{U^\bb}
\ee
is present, without specifying its origin. For the moment $D$ and $b$ are arbitrary positive numbers (the former has the dimension of an energy density, the latter is dimensionless). We will see in section \ref{sec:radial} and in appendix \ref{app:radial} that the concrete value of $b$ is very important to ensure the radial stability of the D-brane, i.e. that the D-brane at the tip is at a local minimum in the radial direction. As we will discuss in the next section, $D$ has to be such that the cosmological constant at the end of inflation is very small. Notice that $V_{up}$ does not depend on the D3-brane position at the tip. 

The potential takes therefore the form
\be 
V(\sigma,\phi)&=&V_{up}+ V_{KKLT}+\Delta V\nonumber\\
&\simeq &\Lambda(\sigma)+B(\sigma)\cos\phi+C(\sigma)\sin^2\phi+\dots\,, \label{33}
\ee
where $\Lambda,\,B$ and $C$ are positive, with the dimension of a $\mathrm{mass}^4$ and can be obtained comparing \eqref{33} with \eqref{ex}; the subleading terms are suppressed at least by a factor $\varepsilon/\mu$.

The potential \eqref{33} depends on two real variables $\sigma$ and $\phi$, but the former gets a much larger mass that the second and we can integrate it out. As we show in appendix \ref{min}, the minimum $\sigma_{cr}(\phi)$ in the $\sigma$ direction has a very mild dependence on $\phi$. In fact, calculating $V(\sigma_{cr}(\phi),\phi)$ this dependence produces terms of the same order as those that we have neglected in \eqref{33}. We are therefore allowed to substitute $\sigma\simeq\sigma_{cr}$ in $\Lambda$, $B$ and $C$ and study the single field potential
\be \label{pot no}
V(\phi)\simeq \Lambda+B\cos\phi+C\sin^2\phi+\dots\,.
\ee
It is important now to determine the relative size of the three coefficients above. To do this we need to know the value of $\sigma_{cr}$. As explained in \cite{Krause:2007jk} and reviewed in appendix \ref{min}, a useful reparameterization of $D$ and $W_0$ is the following
\be 
W_0=-A_0e^{-a\sigma_0}\left[1+\frac13 a (2\sigma_0-k_0) \right]\,,\\
D=\beta\frac13 |A_0|^2a^2e^{-2a\sigma_0} (2\sigma_0-k_0)^{b-1}\,, \label{beta2}
\ee
then, at leading order, the minimum of the volume is
\be 
\sigma_{cr}(\phi)\simeq \sigma_0+\frac{b}{2}\frac{\beta}{a^2\sigma_0}\,+\frac{\varepsilon}{a \mu n}\cos \phi+\frac{\varepsilon^{2/3}}{2c a^3n^2\mu^2 \sigma_0 \gamma}\sin^2\phi \,+\dots
\ee
Substituting $W_0$, $D$ and $\sigma_{cr}$ into $\Lambda$, $B$ and $C$ obtained from \eqref{ex}, and neglecting terms subleading in the $\varepsilon/\mu$ and large volume expansion, we are left with
\be \label{L}
\Lambda&\simeq&\frac{\kq|A_0|^2a^2e^{-2a\sigma_0}}{6\sigma_0}\left[(\beta-1)-\frac{b^2\beta^2}{4(a\sigma_0)^2}+\dots\right]\\
B&\simeq&\frac{\kq|A_0|^2ae^{-2a\sigma_0}\varepsilon}{6n\mu\sigma_0^2} \label{B} \left[(b\beta-3)+\frac{b\beta(14-3b\beta)}{4a\sigma_0} +\dots\right] \\
C&\simeq&\frac{\kq|A_0|^2e^{-2a\sigma_0}\varepsilon^{2/3}}{12cn^2\mu^2\gamma\sigma_0^2}+\dots\,. \label{C}
\ee
We will see in the following that we are interested to the case $\beta\simeq1$ so that the co\-smo\-lo\-gi\-cal constant after inflation is negligible. This implies $(\beta-1)=\mathcal{O}(\sigma_0^{-2})$. Then from the above equations, it is clear that some subtleties would arise in the case $b=3$ because then the factor $(b\beta-3)$ introduces an additional suppression that is not taken into account by the $\varepsilon/\mu$ and large volume expansion. On the other hand, $b=3$ is quite interesting because explicit uplifting mechanisms exist with this scaling \cite{Burgess:2003ic,Saltman:2004sn}. A part for the anti D3-brane at the tip of a warped throat that scales with $b=2$, we are not aware of any other explicit model with $b\neq3$; such scaling could be interpreted\footnote{We thank L. McAllister for suggesting this interpretation \cite{private}. For a few further comments, see appendix \ref{app:up}} as if the physics responsible for the SUSY breaking is localized on a p brane with $p\neq3$. In this case the uplifting would naively scale as $U^{(p-15)/4}$. Other consideration about the role of the uplifting will be given in the next section \ref{app:radial} and in the appendices.

In the following we will separately discuss the case $b=3$ and $b\neq3$. Let us compare $B$ and $C$ to know which is the leading $\phi$ dependent term. The result follows straightforwardly from \eqref{B} and \eqref{C} remembering that $(\beta-1)=\mathcal{O}(\sigma_0^{-2})$: 
\be
C \ll B \left\lbrace  \begin{array}{lcl} \label{BC}
                     (a n c \gamma \mu ) \varepsilon^{1/3} \gg \frac{4}{15}a\sigma_0  & \mathrm{for} & b=3 \\
 		     (a n c \gamma \mu ) \varepsilon^{1/3} \gg 1  & \mathrm{for} & b\neq3\,, 
                   \end{array} \right.
\ee
i.e. in the above regimes $C$ is negligible and $V$ in \eqref{pot no} takes the form of the Natural Inflation potential \cite{Savage:2006tr}. We will see in the next section that in this regime slow roll inflation is impossible, basically because the axion decay constant (that appears when we consider a canonically normalized kinetic term) is always small in Planck units. In section \ref{subsec:nut1} we will study the DBI regime for the Natural Inflation potential. On the other hand, there is an interesting case in which $C$ is non negligible, i.e. the fine tuned case $C=B/2$. As we will see in section \ref{subsec:slowroll}), now the scalar potential supports slow roll inflation. The fine tuning is achieved when the inequality in \eqref{BC} is fulfilled (in the case $b=3$ or $b\neq3$ respectively).

In what follows we estimate $(a n c\gamma\mu)^3$; this will tell us how generic is the Natural Inflation regime $B\gg C$ defined in \eqref{BC} as well as if there is any obstruction from the string theory point of view to achieve the fine tuning $B=2C$. 

From the definition of $a$, it follows $an=2\pi$. The quantity $\mu^{2/3}$ indicates the smallest value $r_{\mu}$ of the radial coordinate $r$ reached by the D7 stack (see figure \ref{fig:mu}). To be able to trust the threshold corrections to the non perturbative superpotential \eqref{conj}, $r_{\mu}$ has to be well inside the warped conifold. In the spirit of \cite{Verlinde:1999fy}, the conifold geometry ceases to be a good description of the compact manifold roughly when the warping $h$ in the 10 dimensional metric
\be\label{metric}
ds_{10}^2=h^{-1/2}ds_4^2+h^{1/2}ds_6^2\,,
\ee
becomes of order one. For the KS solution \cite{Klebanov:2000hb}, $h(r)\simeq(R/r)^4$ where $R^4\equiv(27\pi/4) g_sN \alpha'^2$. Hence an upper bound on $\mu$ is
\be 
\mu < R^{3/2}=\left(\frac{27\pi}{4}g_s N\right)^{3/8}\,.
\ee 
By comparison of the SUGRA kinetic term with the DBI action one obtains $\gamma=\sigma_0T_3/(3\Mp^2)$ \cite{Baumann:2007ah}; expressing it in string units $\gamma=(2\pi)^4g_s/(6\sqrt{\sigma_0})$, where we have used\footnote{We make this choice for the sake of simplicity; it is straightforward to generalize this estimate to the case in which the volume of the Calabi Yau is larger than the volume of the conifold.} $V_6\sim\sigma_0^{3/2}$. 
Putting all the ingredients together we obtain
\be 
(a\gamma\mu n)^3\sim10^{11}\frac{g_s^3(g_sN)^{9/8}}{\sigma_0^{3/2}} (\al)^{-3/4}\,.
\ee
For the choice of parameters $\{g_s=0.1,\,N=10^4,\,\sigma_0=100\}$, we get $(an\gamma c \mu)^3\simeq 3\times10^{8}$. In view of the conditions in \eqref{BC}, e.g. for $b=3$, a value $\varepsilon\gg 6 \times 10^{-8}$ allows to neglect $C$ in the potential. This a value corresponds to a warp factor at the tip $h(r_0)\sim 2 \times 10^{19}$. For such moderately warped throat, we are therefore left with the Natural Inflation potential \cite{Savage:2006tr}. For fine tuned values of $\varepsilon$ saturating this bound, the potential is plotted in figure \ref{fig2} and will be analyzed in section \ref{subsec:hilltop}.

%%%%%%%%%%%%%%%%%%%%%%%%%%%%%%%%%%%%%%%%%%%%%%%%%%%%%%%%%%%%%%%%%%%%%%%%%%%%%%%%%%%%%%%%%%%%%%%%%%%%%%%%%%%%%%%

\subsection{Radial stability}\label{sec:radial}

Until now we have calculated the dependence of the potential on the position of the D3-brane at the tip. Before using this result to produce a stage of inflation, we have to check that the radial direction is not tachyonic, i.e. that if the D3-brane starts at the tip, it will stay there. We leave the details of the calculation to the appendix \ref{app:radial}; here we state and comment the results.

We consider separately the case $C\ll B$, that as we will see can lead to DBI inflation, and the case $B=2C$ that allowing for fine tuning leads to a phenomenologically successful slow roll inflation. The radial stability is determined by the sign of $\partial_{r} V$ evaluated at the tip $r=\varepsilon^{2/3}$.

In both cases, $C\ll B$ and $2C\simeq B$, the tip is a local minimum in the radial direction if the volume scaling of the uplifting $b$ is larger or equal to $3$. On the other hand, for $b<3$ it is typically the case that the radial stability depends on the angular position. In this case, in some part of the tip the radial derivative is positive, in some other negative. To avoid this complication, we consider just the case in which an uplifting with $b\geq3$ is performed.

% The condition $C\ll B $ translated in string parameters for $b>3$ or $b=3$ requires $(an\gamma c\mu)\varepsilon^{1/3}\gg1$ or $(an\gamma c\mu)\varepsilon^{1/3}\gg\sqrt{a\sigma_0}$ respectively. In both cases, this is enough to ensure that the leading term in $\partial_r V$ is always positive.
% The condition $2C=B$ translated in string parameters requires $(an\gamma c\mu)\varepsilon^{1/3}=1$ or $(an\gamma c\mu)\varepsilon^{1/3}=a\sigma_0(4/15)$ for $b>3$ or $b=3$ respectively. In both cases the radial stability is ensured for the whole duration of slow roll inflation, that as we will see in section \ref{subsec:hilltop} takes place next to the top of the potential. Some subtleties might arise at the bottom of the potential and might affect the mechanism of reheating which is out of the focus of the present work.

As an aside, we would like to make two comments about the dependence of the potential on the  radial position of the D3-brane. Both comments refer to models of \textit{radial} brane inflation, which is different from the main focus of the paper which is on \textit{angular} brane inflation. 

The first one is that the uplifting term plays a crucial role for the radial dynamics. In the present setup for example $b\geq3$ is required ($b$ was defined in \eqref{Vup}) to ensure the radial stability at the tip. This suggests the idea of redoing the analysis of KKLMMT \cite{Kachru:2003sx} for a generic uplifting, i.e. $D/U^b$ instead of $D/U^3$. We do this in appendix \ref{app:up} and found that the large inflaton mass problem of KKLMMT can be solved, allowing for fine tuning, in the case $b<2$ or with two uplifting terms of opposite sign. We refer to these terms as \textit{nice upliftings}. In appendix \ref{app:up} we propose some construction to obtain such scaling behavior.

The second comment regards the possibility to cancel the large inflaton mass term using the threshold corrections to the non perturbative superpotential. The result of the investigation of \cite{Krause:2007jk,Baumann:2007ah,Burgess:2006cb}, was that this cannot be achieved for a large range of the radial position because the corrections to $W$ produce only terms\footnote{Or no terms at all, depending on the particular embedding chosen.} like $r,\,r^{3/2},\,r^3,\,\dots$. A large mass for the inflaton is induced by the mixing of the radial position of the D3-brane and the K\"ahler moduli. Consider the scalar potential
\be \label{rad}
V(\sigma,\phi)&\simeq& V_{KKLT}+V_{up}\\
&\simeq & \frac{2\kq A_0ae^{-a\sigma}}{U(r)^2} \Bigg\lbrace A_0e^{-a\sigma}-|W_0|+ \nonumber\\
&& \hspace{1cm}+\frac16 \left[\rho+\bar{\rho}+\gamma\left(k_{\bar{i}}k^{\bar {i} j }k_{j}-k \right)\right]A_0ae^{-a\sigma} \Bigg\rbrace +\frac{D}{U(r)^3}\,.\nonumber
\ee
Far away from the tip but still inside the throat, where the geometry is well described by the singular conifold metric, we have
\be
k(z,\bar z)&=&\left( \sum_{A=1}^{4} |z_A|^2 \right)^{2/3}=r^2 \qquad \Rightarrow \qquad k_{\bar{i}}k^{\bar {i} j }k_{j}=k=r^2 \nonumber \\
U&=&2\sigma-\gamma k(z,\bar z)= 2\sigma-\gamma r^2\,.
\ee
Expanding \eqref{rad} in $r^2\ll\sigma$ one obtains a term $r^2$ that induces a slow roll $\eta$ parameter of order one and prevents slow roll radial inflation. Here we notice that at the tip one has to use the deformed conifold K\"ahler potential
\be \label{Utip}
k(z,\bar z)&=&\tilde{k}_0+ \sum_{A=1}^{4} |z_A|^2 \equiv\tilde{k}_0+ r^3\qquad \Rightarrow \qquad k_{\bar{i}}k^{\bar {i} j }k_{j}=0 \,, \nonumber \\
U&=&2\sigma-\gamma k(z,\bar z)= 2\sigma-\gamma \left(\tilde{k}_0+ r^3\right)\,,
\ee
for some constant $\tilde{k}_0$. By expanding in $r^3\ll\sigma$, we see that near the tip the moduli stabilization induces therefore a cubic term and not a quadratic one as it is the case away from the tip. 
Therefore a cancellation is in principle possible between the threshold corrections to the nonperturbative superpotential and the term induced by the moduli stabilization. This cancellation would happen only close to the tip where the approximated form of the K\"ahler potential \eqref{Utip} is valid, hence it would probably interest only a short range of the radial position.

To summarize, \textit{in the present setup a D3-brane at the tip is at a local minimum in the radial direction as long as $b\geq 3$} (remember that $b$ is the volume scaling of the uplifting).

%%%%%%%%%%%%%%%%%%%%%%%%%%%%%%%%%%%%%%%%%%%%%%%%%%%%%%%%%%%%%%%%%%%%%%%%%%%%%%%%%%%%%%%%%%%%%%%%%%%%%%%%%%%%%%%
\section{Inflationary analysis}\label{sec:inlf}
 
For the inflationary analysis, it is easier to work with a canonically normalized inflaton field:
\be \label{can}
\phi_{can}=\varepsilon^{2/3}\sqrt{T_3 c} \phi \equiv \dd \phi \,.
\ee
Now the potential can be written in the form
\be\label{res}
V(\phi)= \Lambda + B\cos \frac{\phi}{\dd}+C \sin^2\frac{\phi}{\dd}\,,
\ee
where, to keep the notation simple, here and in the following we use again $\phi$ to indicate $\phi_{can}$; $\Lambda$, $B$ and $C$ were defined in \eqref{L}, \eqref{B} and\eqref{C}. For the inflationary analysis, we can think of $\Lambda$, $B$ and $C$ as some constants with the dimension of an energy density that are determined in terms of stringy parameters. 

If we want to be left with a vanishing or very small cosmological constant after inflation, i.e. when $\phi\simeq \pi \dd$, then in \eqref{res} we have to choose the uplifting such that $\Lambda=B$, which from \eqref{L} means $\beta\simeq1$; this leads to the potential
\begin{equation} \label{nat inf}
\framebox[1.3\width]{$V(\phi)= \Lambda \left(1+ \cos\frac{\phi}{\dd} \right)+C\sin^2\frac{\phi}{\dd}$}\,,
\end{equation}
where $0\leq \phi/\dd \leq  \pi$. There are two interesting regimes to analyze: $\Lambda\gg C$ and $2C\simeq \Lambda$. In the first case the potential reduces to the Natural Inflation potential \cite{Savage:2006tr}
\be \label{nat infl pot}
V(\phi)= \Lambda \left(1+ \cos\frac{\phi}{\dd} \right)\,.
\ee
Originally this potential was derived for an axionic field, in which case the parameter $\dd$ is the axion decay constant; we sometimes borrow this terminology. 

Notice that both in the regime $\Lambda=B\gg C$ and $\Lambda=B\simeq 2C$ our expansion in \eqref{ex} is still valid. The terms we neglected in the expansion of $V_{KKLT}$ and $\Delta V$ are suppressed by a factor $\varepsilon/\mu$. These subleading terms can become important close to the minimum of the potential $\phi\simeq \pi d$, where our leading order potential \eqref{nat inf} approaches zero. When the D3-brane reaches that region anyway, inflation is already over, therefore we do not expect these corrections to have any influence on our analysis.

The potential \eqref{nat inf} is the starting point for the phenomenological analysis of this section. As $\phi$ describes the position of a D3-brane (in some angular direction), its kinetic term comes from the DBI action. We will therefore divide our analysis in two parts: first we investigate in section \ref{subsec:slowroll} and \ref{subsec:hilltop} the slow roll regime in which the DBI kinetic term reduces to the canonical one. Second in section \ref{subsec:nut1} we investigate the relativistic regime in which the DBI action is responsible for a behavior very different from the slow roll case.

The search for slow roll is in turn divided in two parts: first in section \ref{subsec:slowroll} we consider \eqref{nat inf} in the regime $\Lambda\gg C$ which reduces to the Natural Inflation potential \eqref{nat infl pot}. A thorough analysis of this potential with a canonical kinetic term has already been performed in \cite{Savage:2006tr}; we review the constraints on the only two parameters $\Lambda$ and $\dd$ imposed by the comparison with WMAP3 data (see figure \eqref{fig1}). Expressing $\Lambda$ and $\dd$ in terms of stringy parameters we show that \textit{our model of Natural Inflation cannot fulfil the constraints}. Basically this is due to the impossibility of obtaining a large axion decay constant in the string theory model. This fact is true also if one tries, considering the collective motion of many D3-branes at the tip, to use the assisted inflation mechanism \cite{Liddle:1998jc}. These difficulties are similar to those found in the case of axionic N-flation \cite{Grimm:2007hs}.

The second part of the slow roll analysis is in section \ref{subsec:hilltop}, where we show that the potential \eqref{nat inf} in the fine tuned regime $\Lambda\simeq 2C$ becomes very flat close to the top (see figure \ref{fig2}). \textit{With this fine tuning, we can have a phenomenologically successful slow roll inflation} with $n_s\simeq0.94$, negligible tensor modes and the scale of inflation $(\Lambda)^{1/4}\sim\dd \times10^{-3}$. 

In section \eqref{subsec:nut1} we come to the DBI analysis. In this case the regime $2C\simeq \Lambda$ does not possess additional interesting features therefore we limit our analysis to the potential \eqref{nat infl pot}. As can be seen from the action in \eqref{action}, the DBI kinetic term introduces a new parameter: $f$ (in string theory it is given by the warp factor times the D3-brane tension). An interesting result is that in the presence of a small speed of sound, contrary to the case of a canonical kinetic term, the CMB data do not require a superplanckian value for $\dd$ anymore. This feature might be relevant for the task of embedding a phenomenologically successful Natural Inflation in string theory. 

\textit{The embedding of DBI Natural Inflation in string theory} that we proposed in the last section \textit{can not satisfy the phenomenological bounds} on $\Lambda$, $a$ and $f$. We prove in section \ref{no go} that this is actually true for any potential at the tip provided that it satisfies the DBI conditions \eqref{cond1} and \eqref{cond2} during the whole duration of inflation. The reason is that if the motion is exclusively relativistic (DBI regime), then inflation at the tip can not last more than few e-foldings. This no go result rules out a large class of potentials and gives us an important criteria to look for a successful model: the potential needs to have, at least somewhere, flat regions such that the slow roll conditions are verified. In the model presented in this paper, these kind of alternating potentials might arise, e.g. considering a general Kuperstein embedding, i.e. a general $\g$ in \eqref{kup}.

%%%%%%%%%%%%%%%%%%%%%%%%%%%%%%%%%%%%%%%%%%%%%%%%%%%%%%%%%%%%%%%%%%%%%%%%%%%%%%%%%%%%%%%%%%%%%%%%%%%%%%%%%%%%%%%

\subsection{Slow roll Natural Inflation}\label{subsec:slowroll}

\begin{figure}
\includegraphics[width=0.9\textwidth,height=0.5\textwidth]{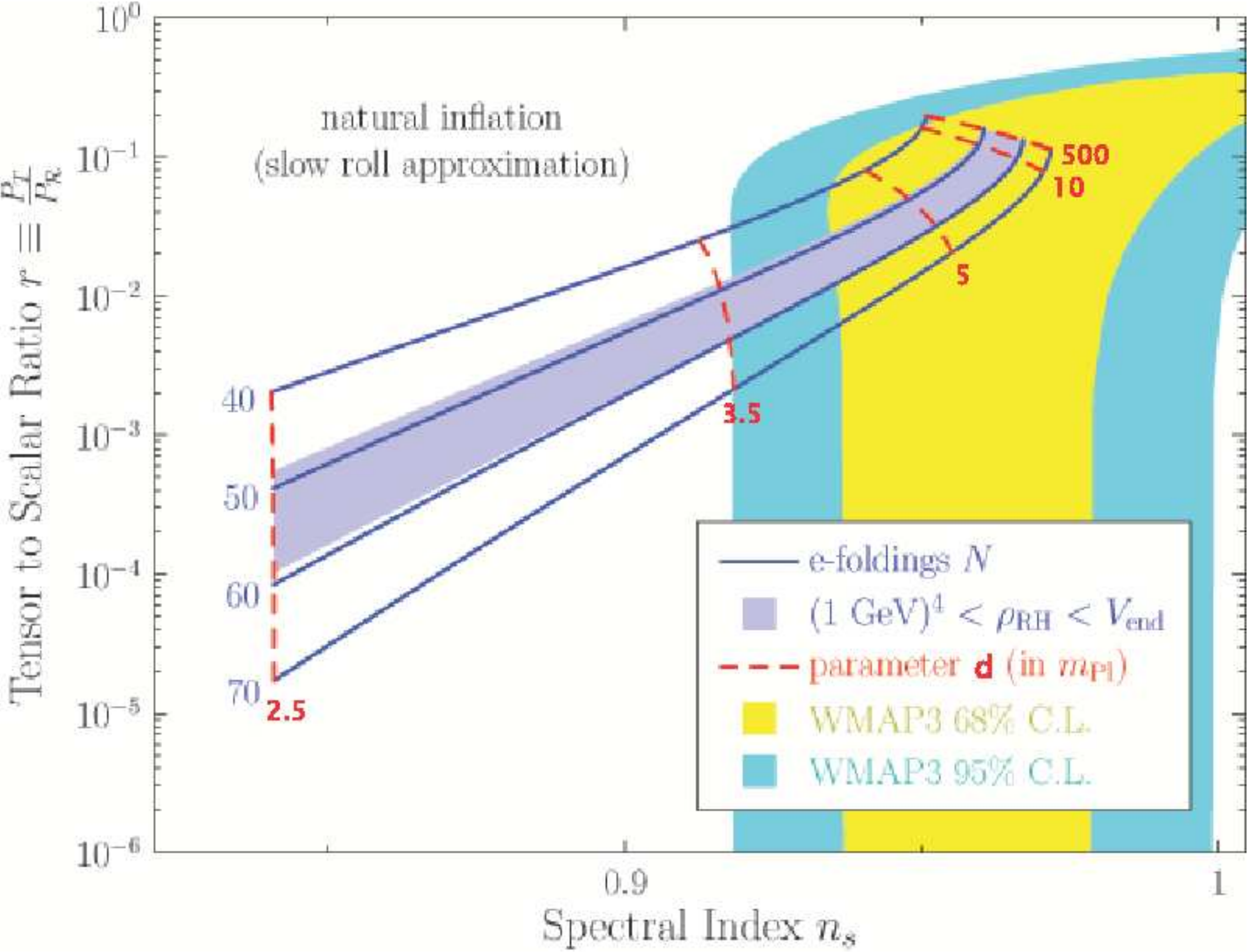}
\caption{WMAP3 data and the predictions of natural inflation are shown in the $n_s-r$ plane. The figure is taken from \cite{Savage:2006tr}.\label{fig1}}
\end{figure}
In this section we consider the inflaton action 
\be \label{nat infl}
S&=&-\int d^4x \, \sqrt{-g}\left[ \frac12 \,  \partial^{\mu}\phi\partial_{\mu}\phi+\Lambda \left( 1+\cos\frac{\phi}{\dd}\right)\right]\,,
\ee
which is the one felt by a D3-brane moving at the tip of a warped deformed conifold when a stack of D7-branes wraps a 4-cycle defined by the supersymmetric Kuperstein embedding \eqref{kup} with constant $\g=\mu$ and the volume is fixed \`{a} la KKLT. This form is valid at leading order in the large volume and $\varepsilon/\mu$ expansion (i.e. when the D7's are not too close to the tip). Also we are in the (large) region of parameter space where we can neglect the $C$ term in \eqref{nat inf}. The effects of this term will be considered in the next section.

The potential \eqref{nat infl} was already studied in \cite{Savage:2006tr}. The model successfully reproduces WMAP3 data \cite{Spergel:2006hy} if\footnote{As it is common in the string cosmology literature, we indicate with $\Mp$ the reduced Planck mass: $8\pi G_{N_{D3}} = \Mp^{-2}$. To compare it with the Planck mass unit one has to multiply the latter with $\sqrt{8\pi}\simeq 5.01$.} $\Lambda\sim m_{GUT}$ and $\dd>0.7\sqrt{8\pi}M_{pl}\simeq 3.5 M_{pl}$, as summarized in figure \ref{fig1}. In this section we show that in the string model it is not possible to have a superplanckian $\dd$. Hence, \textit{as regards slow roll inflation, our model in this regime, $\Lambda\gg C$, is not phenomenologically successful}.

For $\dd\gg\Mp$ the natural inflation approximates the chaotic inflation; in fact the CMB perturbations are produced close to $\phi\simeq \pi \dd$ where the two potentials are indistinguishable. For $\dd<\Mp$ on the other hand, inflation has to start very close to the top of the potential to get enough e-foldings. The problem is that at the top one has $|\eta|\sim (\Mp/\dd)^2\gtrsim 1$, i.e. the slow roll conditions are not satisfied. In the next section we will see how in a different regime ($2C\simeq \Lambda$) from the one considered in this section ($C\ll \Lambda$) and allowing for fine tuning, this problem can be cured.

We now show that in the regime of validity of our stringy model, \textit{the constraint $\dd\gtrsim 3.5 \Mp$ cannot be satisfied}. $\dd=\varepsilon^{2/3}\sqrt{T_3c}$ appears because we work with a canonically normalized inflaton field and it was given in \eqref{can}. The assisted inflation idea \cite{Liddle:1998jc} was used for the first time in string brane inflation in \cite{Becker:2005sg} (for assisted axionic string inflation see \cite{Grimm:2007hs}); it can also be naturally embedded in the present model of inflation at the tip. In fact, if $N_{D3}$ D3-branes are present at the tip, $\dd$ gets multiplied by a factor $\sqrt{N_{D3}}$, which goes in the right direction to reproduce the CMB data. The details of how to obtain this result are left to the appendix \eqref{app:many}. 

We now look for a set of stringy parameters that lead to $\sqrt{N_{D3}}\dd>3.5 M_{pl}$. Several constraints are imposed by consistency. One comes from considering the backreaction of the $N_{D3}$ D3-branes on the deformed conifold geometry. For the warped deformed conifold to be a solution of Einstein's equations, the following 3-form fluxes have to be present
\be 
\frac{1}{(2\pi)^2\alpha'}\int_{A}F_3=M\,,\qquad \frac{1}{(2\pi)^2\alpha'}\int_{B}H_3=-K\,.
\ee
This solution possesses $K$ times $M$ D3 charge. As long as the number $N_{D3}$ of D3-branes responsible for inflation is much smaller than the background D3 charge $KM$, we can neglect their backreaction on the geometry at leading order.

The requirement $d\sqrt{N_{D3}}>3.5 M_{pl}$ leads to
\be \label{cond}
N_{D3}>15.8 \,\varepsilon^{-4/3} \frac{M_{pl}^2}{T_3}\,.
\ee
The D3-brane tension is usually given by 
\be
T_3&=&\frac{\al^{-2}}{(2\pi)^3g_s}\,.
\ee
Here we have to consider the effect of the warping, i.e. that the string scale at the tip of the throat is $M_s a_0$ with $a_0=\varepsilon^{2/3}/(g_s M\al)^{1/2}$ the warp factor at the tip. Therefore we have
\be \label{T33}
\textrm{at the tip:   }\qquad \frac{T_3}{M^4_{pl}}=\frac{(2\pi)^{11}g^3_s}{4V_6^2}a_0^4 =\frac{(2\pi)^{11}g_s}{4V_6^2M^2} \frac{\varepsilon^{8/3}}{\al^{2}}
\ee
The volume (in string units) can be written as sum of the conifold plus the rest of the CY, so an obvious lower bound is \cite{DeWolfe:2004qx}
\be 
V_6 > V_{conifold}\simeq\left(\frac{2\pi}{3}\right)^3\left(\frac{27\pi}{4}g_s MK\right)^{(3/2)}\,.
\ee
Substituting this and \eqref{T33} in \eqref{cond} one obtains
\be 
N_{D3}>K^{3/2}(g_sM)^{5/2}\frac{\al^{3}}{\varepsilon^{4}} 18\,.
\ee
For the supergravity approximation to be valid, the radius of the tip has to be large in string units therefore $(g_sM)^{1/2}\gg1$. Even in the most favorable case of very shallow throats, a superplanckian value of $\dd$ requires $N_{D3}\gg KM$ which makes it inconsistent to neglect the backreaction.

Even if one could take into account the backreaction of the inflating D3-branes, we expect that the situation would not improve much. To see this suppose that $N_{D3}\gg MK$; then the AdS radius scales as $R^4\sim N_{D3}$ and the volume of the conifold as $V_6\sim N_{D3}^{3/2}$. Again \textit{it seems to be impossible to fulfil \eqref{cond}, i.e. to get $\dd$ of order one or larger}.
This difficulty is very similar to the one found in axionic natural assisted inflation \cite{Grimm:2007hs}, where no controllable string compactification has been found with $\dd>\Mp$.

%%%%%%%%%%%%%%%%%%%%%%%%%%%%%%%%%%%%%%%%%%%%%%%%%%%%%%%%%%%%%%%%%%%%%%%%%%%%%%%%%%%%%%%%%%%%%%%%%%%%%%%%%%%%%%%

\subsection{Slow roll hilltop inflation}\label{subsec:hilltop}

In this section we consider the potential \eqref{nat inf} for a single D3-brane at the tip, in the regime in which $C$ is non negligible. We find that it is possible to fine tune the string parameters such that the potential at the top becomes very flat. This hilltop model \cite{Kohri:2007gq} \textit{gives rise to a prolonged stage of inflation and is perfectly compatible with WMAP3}.

Let us expand the potential close to the top where $\phi\ll\dd$. The result is
\be \label{mass}
V\simeq 2 \Lambda+\frac{1}{2\dd^2}(2C-\Lambda)\phi^2+\frac{1}{24\dd^4}(\Lambda-8C)\phi^4+\mathcal{O}\left(\frac{\phi^6}{\dd^6}\right)\,.
\ee
For a generic $C\sim \Lambda$ the slow roll parameter $\eta=\Mp^2V''/V$ is nowhere small as we mentioned in the last section. But if we fine tune $C=\Lambda/2$, close enough to the top of the potential\footnote{As usual, $\esr$ is the slow roll parameter $\Mp^2(V'/V)^2/2$} $\esr\ll\eta\ll 1 $. In terms of stringy parameters, this fine tuning can be achieved, e.g. in the case $b=3$, varying the fluxes such that $\varepsilon=(2\pi\mu\gamma c)^{-3}4a\sigma_0/15)$ (that we estimated in \eqref{BC}). Notice that this can be achieved independently from the fine tuning of the cosmological constant left after inflation $\Lambda=B$ that is obtained varying the uplifting ($\beta$).
\begin{figure}
\centering
\includegraphics[width=0.5\textwidth,height=0.4\textwidth]{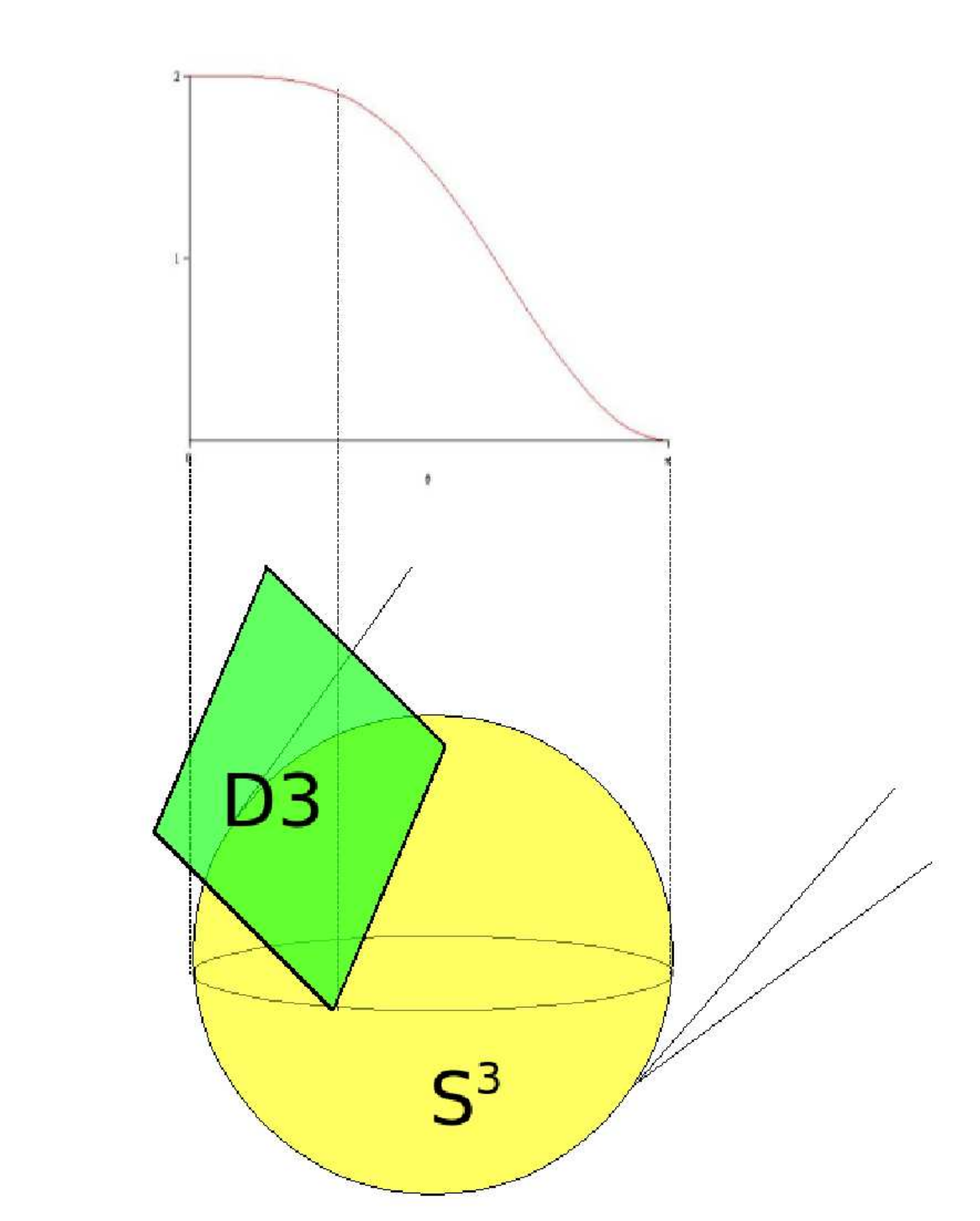}
\includegraphics[width=0.4\textwidth,height=0.3\textwidth]{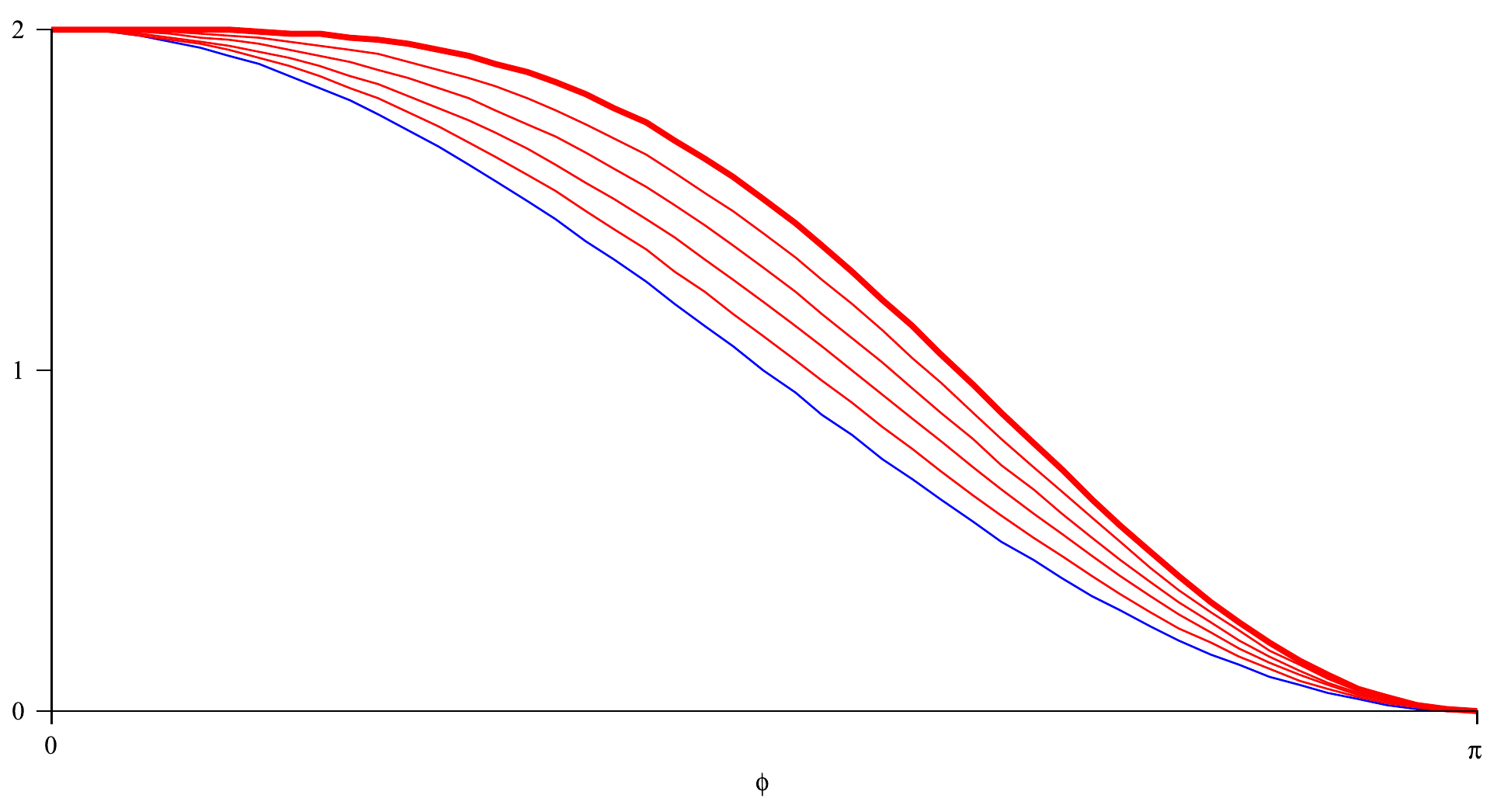}
\caption{On the left: a cartoon of the hilltop inflation model. On the right: we show explicitly how the fine tune works. The potential \eqref{nat inf} is plotted for $C/\Lambda=\{0.5,\,0.4,\,0.3,\,0.2,\,0.1,\,0\}$. The thick line corresponds to $C=\Lambda/2$, i.e. a perfect cancellation of the mass term in \eqref{mass}. The lowest (blue) line corresponds to $C=0$, i.e. the natural inflation potential \eqref{nat infl pot}.\label{fig2}}
\end{figure}

As the cancellation $2C-\Lambda$ becomes more and more precise, the potential at the top becomes flatter and flatter as can be seen from figure \ref{fig2}. For simplicity, in the following we study the case in which the cancellation is precise enough to neglect the mass term\footnote{The case in which $C\sim\Lambda/2$ and both the quadratic and quartic terms are of the same order could also be considered, see for example \cite{Dine:1997kf}.}, such that the potential is well approximated by
\be \label{quartic}
V\simeq 2\Lambda\left(1-\frac{1}{16\dd^4}\phi^4\right)\,.
\ee
The COBE normalization imposes the constraint $\Lambda^{1/4}\sim\dd\times10^{-3}$ \cite{Liddle:2000cg}. As we discussed in the last section, typically in our string theory model $\dd<\Mp$; this implies that the scale of inflation can be somewhat smaller than the GUT scale. To estimate the number of e-foldings $N_e$ before the end of inflation when the scales of the CMB perturbations left the Hubble horizon we can use the formula
\be
N_e\simeq 60-\ln \frac{10^{-16} \,\mathrm{Gev}}{V^{1/4}}\simeq 58.4+\ln \frac{\dd}{\Mp}\,.
\ee
For example, $\dd\sim2\times 10^{-4}\Mp$ gives $N_e\simeq50$. For the potential\footnote{Generically for a potential $V\sim 1-\mu\phi^{p}$ with $p\geq3$ we have $\eta=-(p-1)/[(p-2)N_e]$ \cite{Liddle:2000cg}} \eqref{quartic} $\eta\simeq2/(3N_e)$ and, as is typical for small field models $\esr\ll\eta$. Therefore we have a prediction for the scalar spectral index $n_s\simeq1+2\eta\sim0.94$ in good agreement with WMAP3 \cite{Spergel:2006hy}.

To summarize, our model of inflation at the tip gives the potential in \eqref{quartic} provided that\footnote{The problem of fine tuning the uplifting to get the right cosmological constant is somehow a different problem from the one of embedding inflation in string theory and we do not address it here.} we can fine tune the warp factor at the tip with enough precision to make the mass term negligible. Then the potential \eqref{quartic} gives prediction in good agreement with the experiment if $\Lambda\sim d^4 \times 10^{-12}$. In a numerical investigation we found explicit values that can satisfy this requirement. We conclude that, \textit{allowing for fine tuning, inflation at the tip can provide a phenomenologically viable prolonged stage of slow roll inflation.}

%%%%%%%%%%%%%%%%%%%%%%%%%%%%%%%%%%%%%%%%%%%%%%%%%%%%%%%%%%%%%%%%%%%%%%%%%%%%%%%%%%%%%%%%%%%%%%%%%%%%%%%%%%%%%%%

\subsection{DBI Natural Inflation}\label{subsec:nut1}

In the last two sections, we looked for slow roll inflation and we were therefore allowed to approximate the kinetic term of the D3-brane with a canonical one. In this section, we investigate the possibility of obtaining successful DBI inflation, i.e. with a potential that is too steep to fulfil the slow roll conditions.

Consider the action
\be \label{action}
S=-\int d^4x \sqrt{-g}\left(f^{-1}\sqrt{1+f\partial^{\mu}\phi\partial_{\mu}\phi} +V(\phi)-f^{-1}\right)\,
\ee
where the potential $V(\phi)$ is the one in \eqref{nat inf} and $f$ is a constant with the dimension of a $\mathrm{mass}^{-4}$. We consider the general regime in which the $C$ term in \eqref{nat inf} is very small and can be neglected leaving the Natural Inflation potential in \eqref{nat infl pot}. Actually, as we will see in the next section, as long as we are interested just in DBI inflation, the precise shape of the potential is not important.

The effective action \eqref{action} has three parameters: $\dd,f$ and $\Lambda$. We focus on the regime $\dd\ll\Mp$, which is what happens generically in string theory. We study how these three parameters are constraint by WMAP3 data. The result is that \textit{the phenomenological constraint for DBI inflation can not be fulfilled in the string theory model}. In section \ref{no go} we generalized this negative result to any potential provided that it is nowhere slow roll flat. As we describe in \ref{no go2}, what would work is a potential that alternates slow roll flat regions with steep ones. To obtain such a potential in our model of inflation at the tip, we should consider more generic embeddings than the simple one with constant $\g$ that leads to the potential \eqref{nat inf}.

En passant we obtain an interesting result: as we reviewed in section \ref{subsec:slowroll}, the consistency with CMB data requires that the axion decay constant $\dd$ of Natural inflation with a canonical kinetic term takes superplanckian values. We will see that in the presence of a DBI kinetic term, this constraint is relax and viable (DBI) inflationary model occurs for $\dd$ as small as $0.04\Mp$\footnote{This is a statement from the effective action point of view, just considering the action \eqref{action}. As we said in the string theory setup we can not arbitrarily vary the parameters and therefore the DBI Natural Inflation can not be realized.}. 

%%%%%%%%%%%%%%%%%%%%%%%%%%%%%%%%%%%%%%%%%%%%%%%%%%%%%%%%%%%%%%%%%%%%%%%%%%%%%%%%%%%%%%%%%%%%%%%%%%%%%%%%%%%%

\subsubsection{DBI in a nutshell}\label{subsec:nut}

For completeness, in this section we briefly review the important formulae in the analysis of DBI inflation. 

The idea of DBI inflation \cite{Alishahiha:2004eh} is that, for a D-brane moving in a warped space, the maximal allowed speed can be considerably smaller than the speed of light. This allows to obtain inflation even when the inflaton potential is not as flat as required by the standard slow roll conditions. In the following we collect the relevant steps for a DBI inflation analysis, and refer the reader to the original papers \cite{Alishahiha:2004eh} for further discussions.

Given the action \eqref{action}, the energy density $\rho$ and pressure $p$ in the perfect fluid approximation are
\be 
\rho&=&\frac{\gamma}{f}+(V-f^{-1})\,,\\
p&=&\frac{1}{f\gamma}-V-f^{-1}\,,
\ee
where 
\be \label{gamma}
\gamma=\frac1{\sqrt{1-f\dot{\phi}^2}}
\ee
in analogy with the usual notation of special relativity\footnote{Clearly $\gamma$ here has nothing to do with the one in section \eqref{ST}, e.g. in \eqref{K}. As we separate the inflationary from the string theory analysis, we hope no confusion will arise.}. This implies a speed limit $\dot{\phi}\leq f^{-1/2}$. 

Using the Hamilton-Jacobi formalism, where $\phi$ plays the role of the time variable, the equivalents of the Friedmann equation and of the equation of motion for $\phi$ are
\be 
\dot{\phi}&=&-\frac{2H'}{\sqrt{\Mp^{-4}+4fH^{'2}}}\,, \label{phidot}\\
H^2&=&\frac{\rho}{3\Mp^2}\,.
\ee
Defining 
\be 
\epsilon_{DBI}= \frac{2\Mp^2}{\gamma}\left(\frac{H'}{H}\right)^2\,,
\ee
we have that inflation lasts as long as $\epsilon_{DBI}<1$. In analogy with the slow roll case, we have two DBI conditions  \cite{Alishahiha:2004eh}
\be 
\frac{V^{3/2}}{|V'|\Mp}\sqrt{3f} \gg 1 & \qquad \Rightarrow \qquad & H^2\simeq\frac{V}{3\Mp^2}\,,\label{cond1}\\
\frac{V^{'2}}{3V}f\Mp^2\gg 1  & \qquad \Rightarrow \qquad & \gamma \gg 1\, \label{cond2}
\ee
that guarantee respectively that the energy density is dominated by the potential term and that the motion is relativistic. 

%%%%%%%%%%%%%%%%%%%%%%%%%%%%%%%%%%%%%%%%%%%%%%%%%%%%%%%%%%%%%%%%%%%%%%%%%%%%%%%%%%%%%%%%%%%%%%%%%%%%%%%%%%%%%%%

\subsubsection{DBI Natural Inflation}

In this section we study the DBI regime of a brane moving along the tip under the Natural Inflation potential \eqref{nat infl pot}. We are in the (large) region of parameter space where the term $C$ in \eqref{nat inf} is very small and can be neglected. The conclusion is that \textit{this model of exclusively DBI inflation at the tip can not give enough e-foldings}. As we will see in section \ref{no go}, this is true for any potential provided that it does not possess any slow roll region. 

For the Natural Inflation potential \eqref{nat infl pot}, as long as the conditions \eqref{cond1} and \eqref{cond2} are fullfilled, we have 
\be\label{H}
H=\frac{\sqrt{\Lambda(1+\cos\frac{\phi}{\dd}) } }{\sqrt{3}\Mp}\,.
\ee
Using \eqref{phidot} we can easily obtain $\dot{\phi}$. The number of e-foldings is then given by
\be \label{efoldings}
N_e=\int_{t_i}^{t_f}Hdt=\int_{\phi_i}^{\phi_f}\frac{H}{\dot{\phi}}d\phi\,.
\ee
where the suffixes $i$ and $f$ refer to the beginning and the end of inflation. 

\begin{figure}
\centering
\includegraphics[width=0.4\textwidth,height=0.3\textwidth]{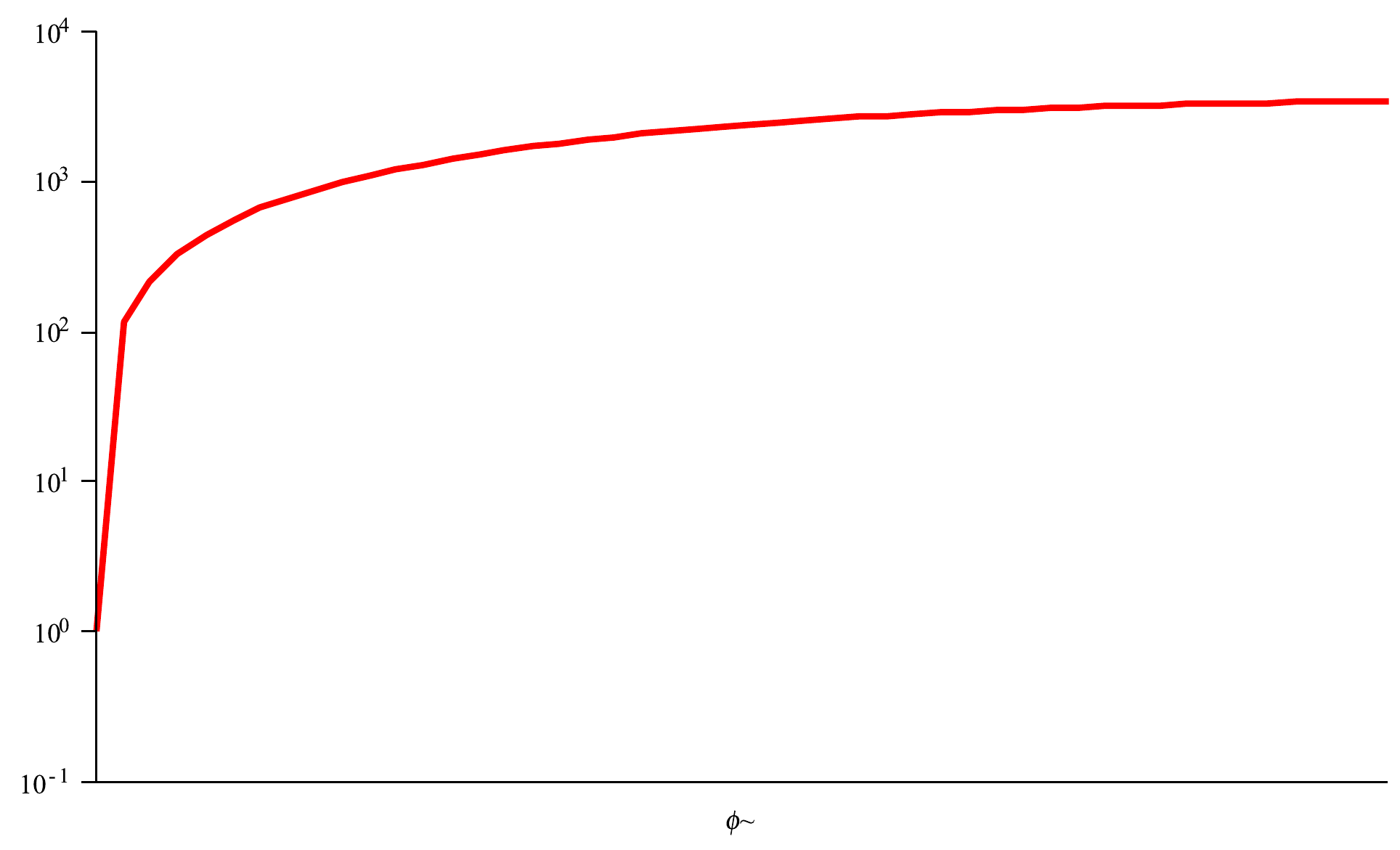} \hspace{0.3cm}
\includegraphics[width=0.4\textwidth,height=0.3\textwidth]{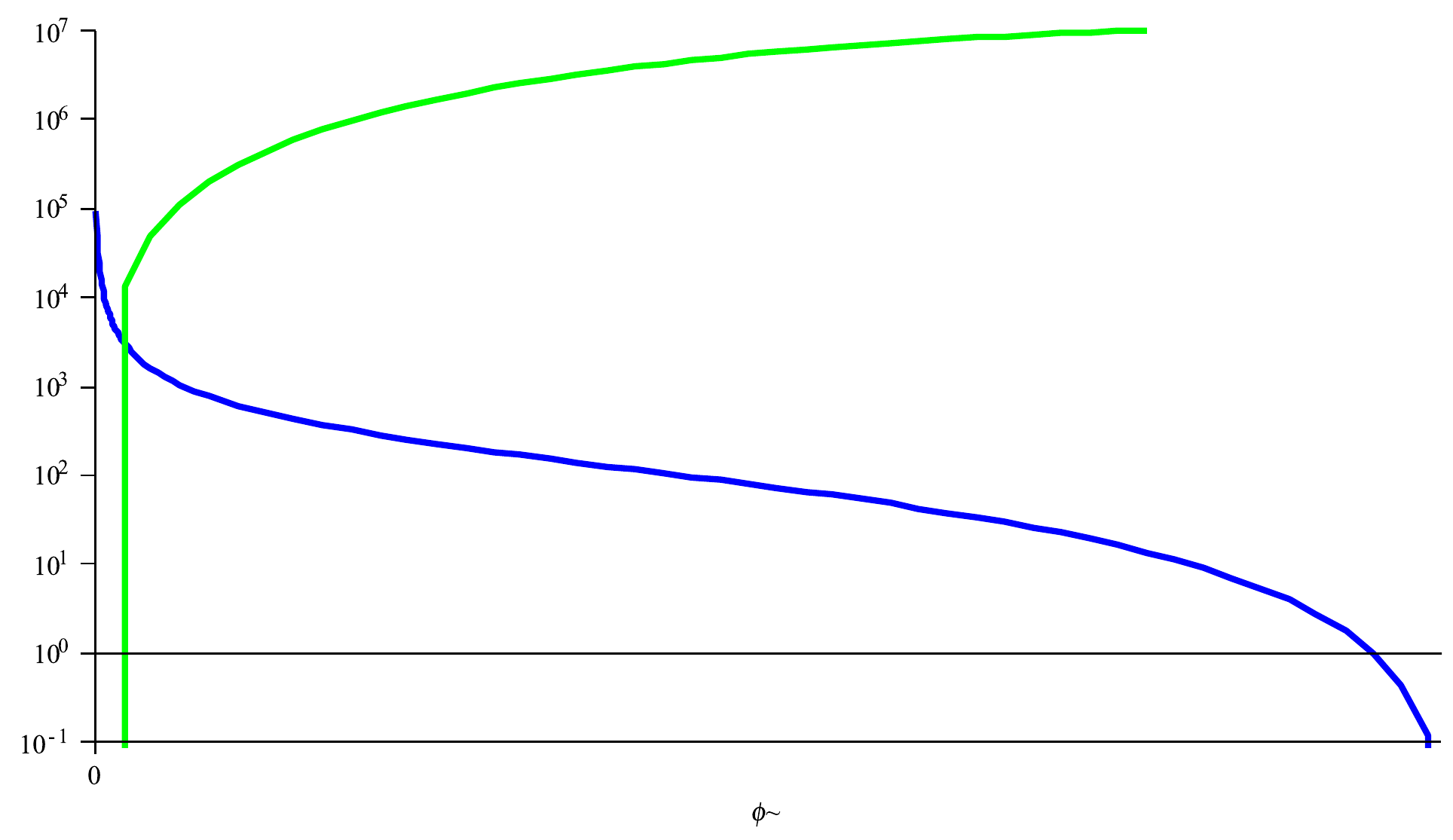}
\caption{On the left: $\gamma$ is plotted for the Natural Inflation potential; it grows monotonically from 1 to $(2f\Lambda/3)^{1/2}\Mp/\dd$.\label{fig:conds} On the right: the plot shows that the DBI condition \eqref{cond1} and \eqref{cond2} for the Natural Inflation potential \eqref{nat infl pot} with $d\ll\Mp$ are satisfied everywhere except for the two extremal regions, $\phi\simeq0$ and $\phi\simeq\dd\pi$.}
\end{figure}

Close to the top of the potential the inflaton moves non relativistically (as found also in \cite{Kecskemeti:2006cg}); this regime is uniteresting for us because, as we said, in that region $\eta\sim-(\Mp/\dd)^2$ which for small $\dd/\Mp$ gives a very red spectrum. Therefore we look\footnote{Here we do not address the problem of initial conditions for the inflaton. Recently in \cite{Underwood:2008dh} it has been shown that in the case of DBI inflation this problem can be much less dramatic than for the canonical case.} further away from the top where the motion becomes relativistic, $\gamma\gg 1$. The importance of relativistic effects can be estimated evaluating $\gamma$. Substituting the solution \eqref{phidot} for $\dot{\phi}$ in the definition \eqref{gamma}, we get
\be\label{gamma sol}
\gamma=\frac1{\sqrt{1-f\dot{\phi}^2}}=\frac{ \Mp}{\dd\sqrt{3}}\sqrt{f\Lambda (1-\cos \frac\phi \dd)+3\dd^2}   \,,
\ee
that is plotted in the left part of figure \eqref{fig:conds}. $\gamma$ grows monotonically from one (non relativistic motion) to $\Mp\sqrt{f\Lambda}/\dd$. 

To avoid unobserved large non Gaussianities in the CMB, the perturbations have to be produced when $\gamma \lesssim 22$ \cite{Alishahiha:2004eh}. This happens close to the top of the potential where we can expand in $\phi/\dd$
\be \label{phi cmb}
\gamma \simeq \frac{\Mp \phi}{\sqrt{6}\dd^2}\sqrt{f\Lambda} \qquad \Rightarrow \qquad \frac{\phi_{CMB}}{ \dd}<\frac {\phi(\gamma=22)}{ \dd}\simeq  22 \sqrt6 \frac{ \dd}{ \Mp \sqrt {f \Lambda}}\,.
\ee
Inflation ends when $\epsilon_{DBI}\gtrsim 1$, or equivalently when the condition \eqref{cond1} is no longer fulfilled and the kinetic energy becomes comparable to the potential energy. From figure \eqref{fig:conds} we see that $\epsilon_{DBI}$ becomes of order one for $\phi\rightarrow \pi\dd$. Expanding around $\phi=\pi\dd$ we get
\be 
\frac{V^{3/2}}{|V'|\Mp}\sqrt{3f}=-\frac{ \dd \sqrt{3 \Lambda f}  \left(1+\cos \frac\phi \dd\right)^{3/2} }{\Mp \sin \frac\phi d }\simeq\sqrt{3\Lambda f}\frac{(\phi-\pi\dd)^2}{2\sqrt{2}\Mp \dd}\nonumber \,,
\ee
which leads to the following analytical estimate (accurate for large $d \sqrt{\Lambda f} /\Mp$) for $\phi_f$
\be\label{phi f}
\sqrt{3\Lambda f}\frac{(\phi_f-\pi\dd)^2}{2\sqrt{2}\Mp \dd}=1 \qquad \Rightarrow \qquad \phi_f=\pi\dd\left[1-\frac{2^{3/4}\sqrt{\Mp}}{\pi\sqrt{\dd}(3f\Lambda)^{1/4}}\right]\,.
\ee
Now that we know where the perturbations have to be produced, $\phi=\phi_{CMB}$, and where inflation ends, $\phi=\phi_f$, we can impose that the number of e-foldings\footnote{As we already mentioned, the number of e-foldings $N_{CMB}$ before the end of inflation when the CMB perturbations are produced can be different from $60$ by some $30\%$, e.g. depending on the reheating temperature, etc\dots. For concreteness we take $N_{CMB}=60$, but another choice would not alter the conclusions of our analysis.} in between is approximatively 60. From \eqref{efoldings}
\be\label{efolds}
60=N_e=\int_{\phi_{CMB}}^{\phi_f}\frac{H}{\dot{\phi}}d\phi \sim \frac \dd \Mp \sqrt{f \Lambda}\,.
\ee
A consistency check is that for small $\dd/\Mp$ if we require $N_e\sim\dd\sqrt{f\Lambda}/\Mp\sim 60$, then the expansions we used in \eqref{phi cmb} and \eqref{phi f} are accurate.

The COBE normalization for the amplitude of the scalar perturbations gives us the constraint
\be 
2\times 10^{-5}\simeq\delta_H=\frac{H^2\sqrt{f}}{5\pi} \qquad \Rightarrow \qquad \Lambda \sqrt{f}\simeq\frac{\Mp^2}{2000}\,.
\ee

The tensor modes are negligible: \mbox{$\Delta\phi\sim \pi \dd<\Mp$}. An upper bound is obtained estimating $r$ at $\phi(\gamma=20)/\Mp\simeq(\dd/\Mp)^2$ obtained in \eqref{phi cmb}
\be 
r\leq\frac{16 \varepsilon}{\gamma}\bigg|_{\phi(\gamma=20)}\simeq\frac{12}{f\Lambda}\,.
\ee

\begin{figure}\label{fig:ns}
\centering
\includegraphics[width=0.32\textwidth,height=0.35\textwidth]{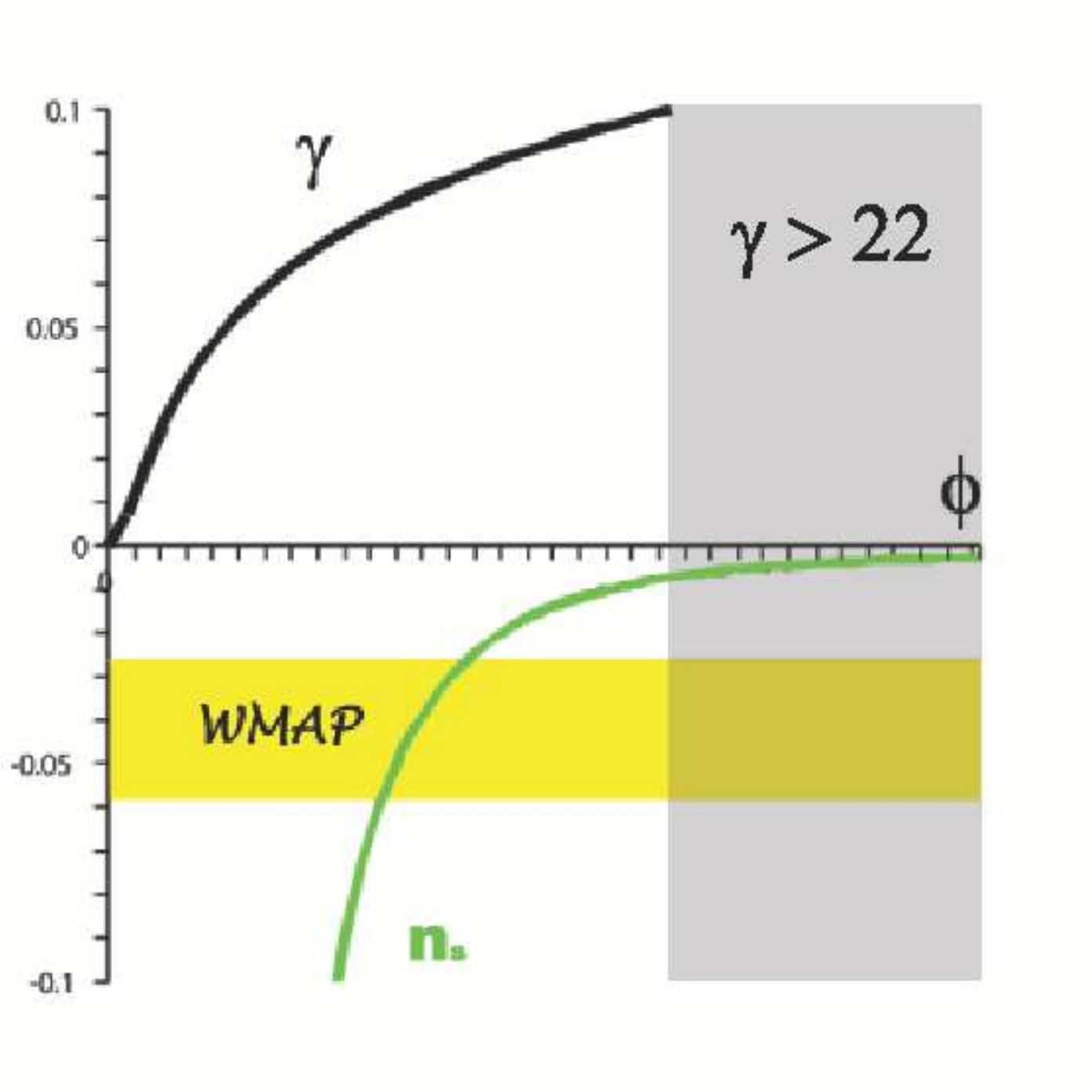}
\includegraphics[width=0.32\textwidth,height=0.35\textwidth]{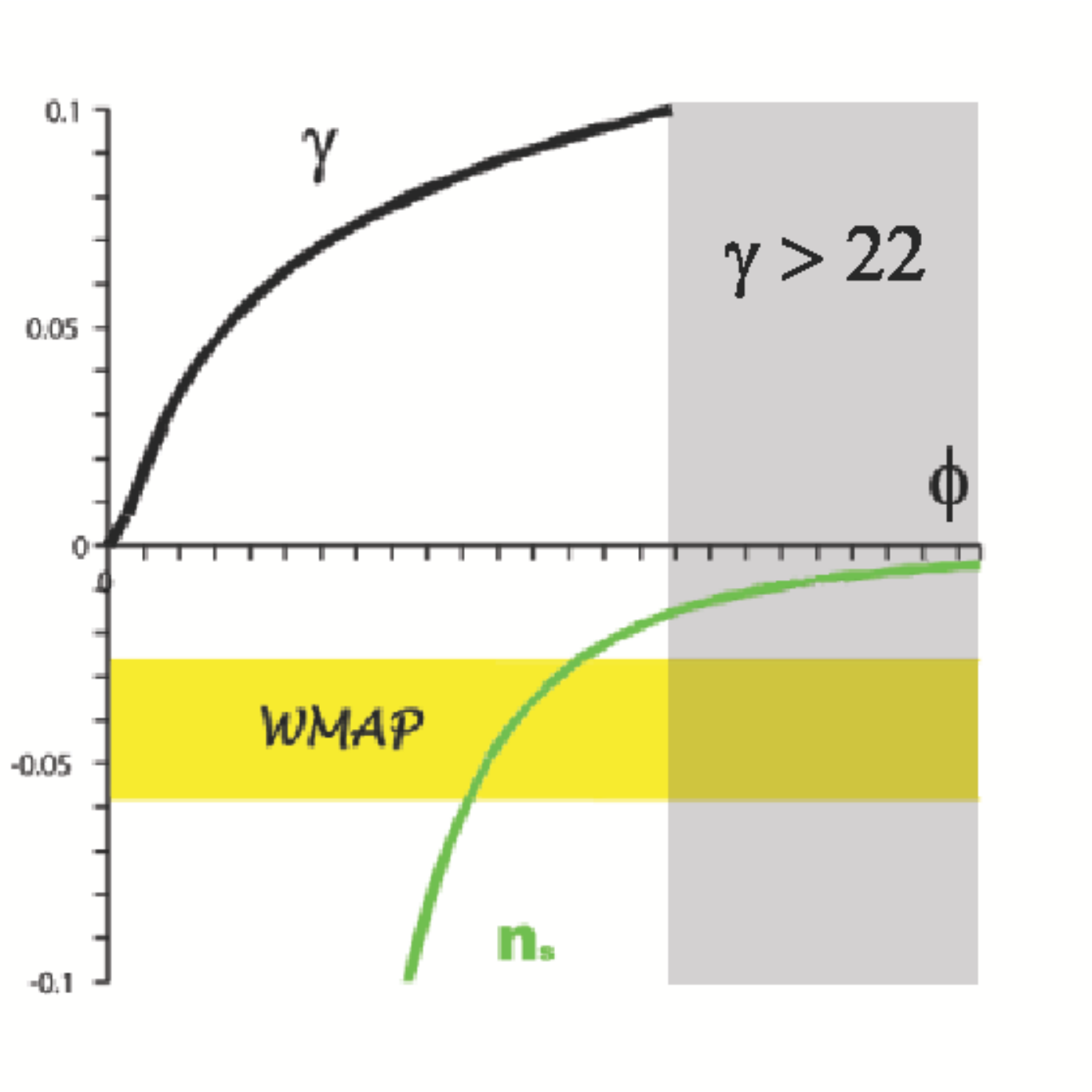}
\includegraphics[width=0.32\textwidth,height=0.35\textwidth]{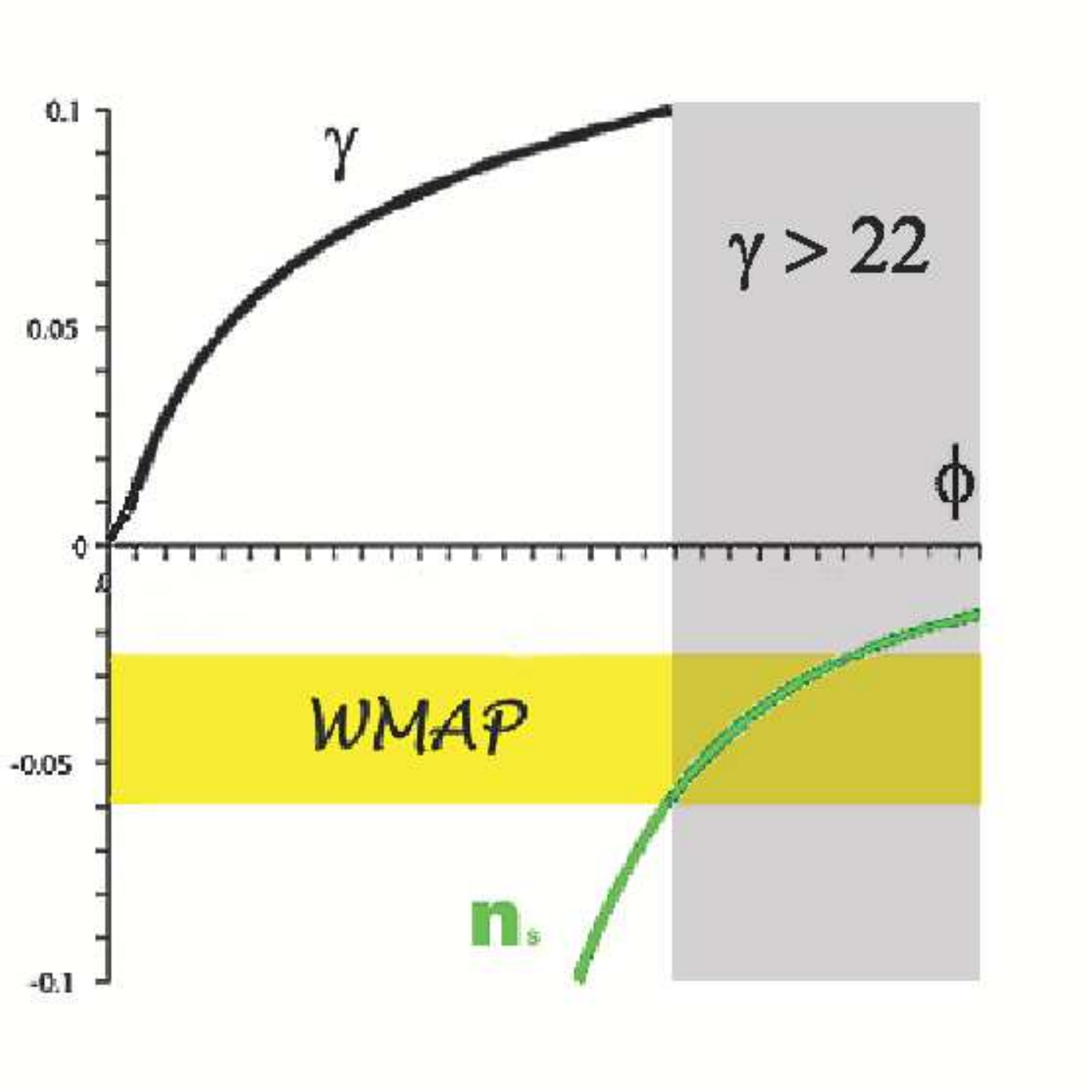}
\caption{The scalar spectral index $n_s$ and $\gamma$ are plotted for three different values $\dd=0.12,\,0.08,\,0.04$. For $\dd<0.04$ the requirements $n_s=0.958\pm0.016$ and $\gamma<22$ cannot be satisfied at the same time.}
\end{figure}

\begin{figure}
\centering
\includegraphics[width=0.45\textwidth]{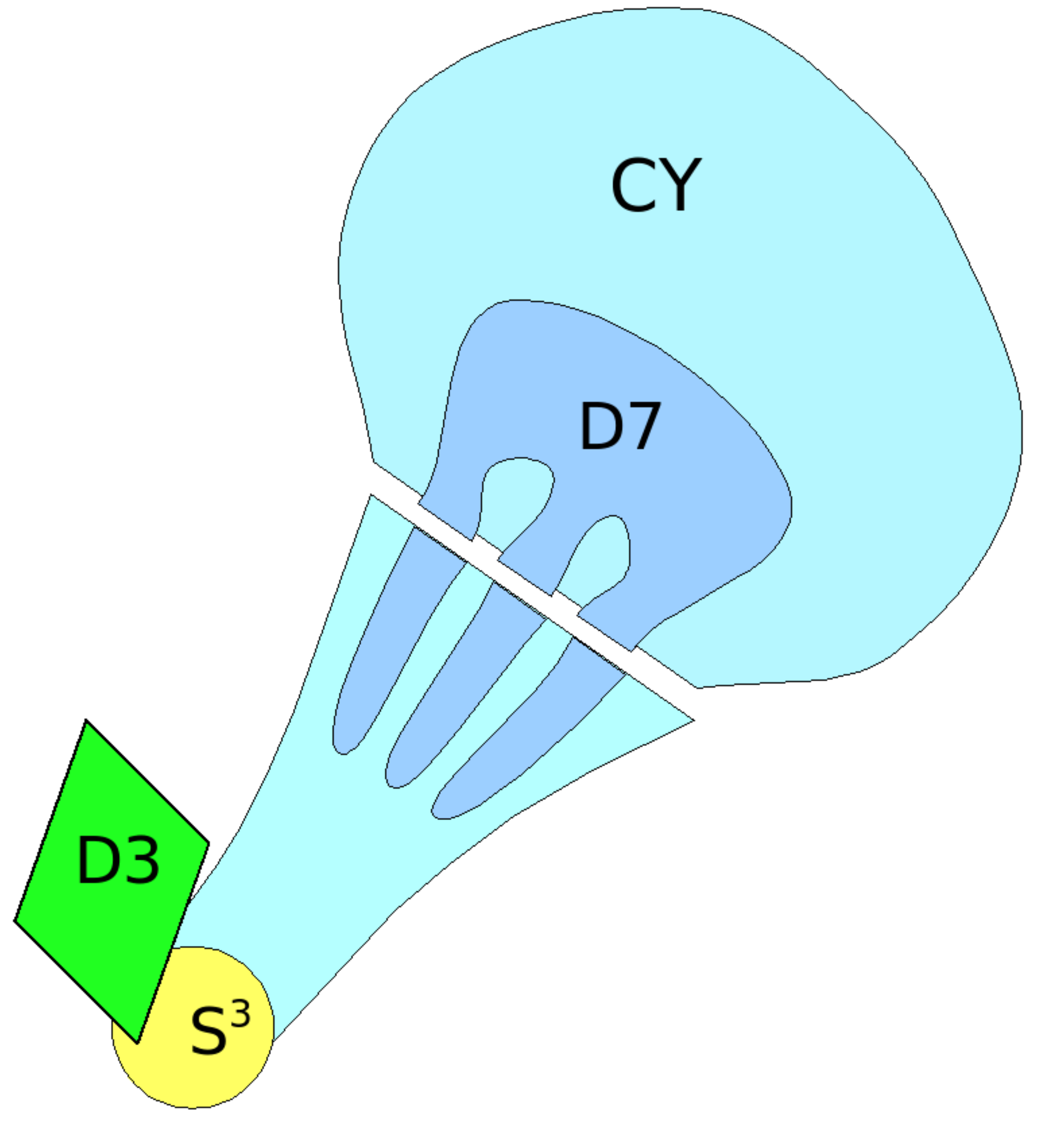}
\includegraphics[width=0.45\textwidth]{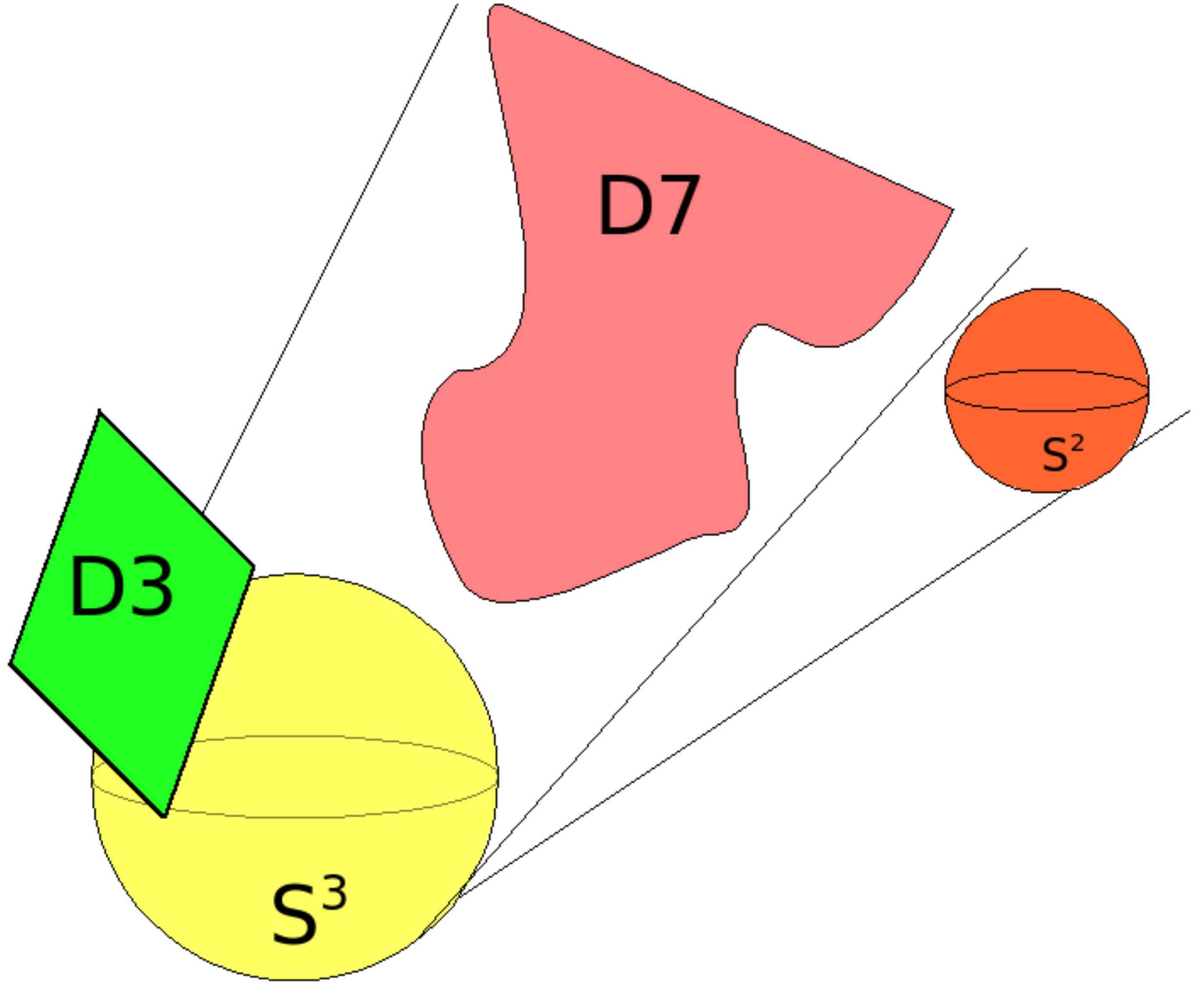}
\caption{Different possibilities for inflation at the tip: on the left we draw the ``threfold'' D7 embedding in the warped throat; after gluing to a compact Calabi Yau the three 4-cycles are just parts of a single globally defined 4-cycle. On the right a possible generalization is depicted where instead of the simplest Kuperstein embedding, one consider a less symmetric one for the D7-branes.\label{recy}}
\end{figure}

Summarizing, we can use the COBE normalization and the requirement of 60 e-foldings to express two of the three parameters of the model in terms of the third one:
\be
\left\lbrace d,f,\Lambda\right\rbrace &\qquad\longrightarrow\qquad & \left\lbrace d,\,f=\frac{72^2\,10^{10}}{d^4},\,\Lambda=\frac{\dd^2\Mp^2}{12^2 10^8} \right\rbrace\,.
\ee
A constraint on $\dd$ comes from the spectral index of the scalar perturbations $n_s$. The analytical expression for $n_s$ is not so illuminating; we plot $n_s$ in figure \eqref{fig:ns} for various values of $\dd$. After imposing a precise value for the scalar spectral index, e.g. $n_s=0.958$, $\gamma$ just depends on $d$; hence detecting some non Gaussianities would determine $d$ and completely fix the parameters of the model. As it is clear from figure \eqref{fig:ns}, the two constraints $n_s=0.958\pm 0.016$ and $\gamma<22$ lead to a lower bound on $\dd$. This can be estimated numerically as $\dd>0.04 \Mp $.

Thus one of the effects of the DBI kinetic term is to relax the bound $\dd>3.5\Mp$ obtained in \cite{Kecskemeti:2006cg} to $\dd>0.04\Mp$. This means that, provided one can arbitrarily choose $\Lambda$ and $f$, superplanckian values of $\dd$ are not required anymore\footnote{We would like to stress that this statement is meant from an effective action point of view. In a concrete (e.g. string theory) model one should carefully consider the setup: the subplanckian (effective) parameter $\dd$ might arise from a superplanckian string parameter redshifted by the large warping.}.

To finish this section we want to address the question whether it is possible to reproduce CMB data within the stringy model of section \eqref{ST}. The answer is that \textit{this embedding of DBI inflation in string theory is not phenomenologically viable.} The reason is the following: from our discussion in section \ref{sec:kup}, we know that generically $\dd<\Mp$; from \eqref{efolds} it is clear that if we want to get 60 e-foldings we need $f\Lambda\gg1$. As we show in detail in the next section, this implies that the energy density during inflation is bigger than the warped string scale and our supergravity approximation is inconsistent. As we will discuss in section \ref{no go2} this problem can be overcome if one considers potentials that possesses both regions satisfying slow roll and DBI conditions \eqref{cond1},\eqref{cond2}.

%%%%%%%%%%%%%%%%%%%%%%%%%%%%%%%%%%%%%%%%%%%%%%%%%%%%%%%%%%%%%%%%%%%%%%%%%%%%%%%%%%%%%%%%%%%%%%%%%%%%%%%%%%%%%%%

\section{A no-go result and how to evade it}\label{no go}

In this section we derive a no-go result for string theory DBI inflation along the tip, i.e. with constant warp factor. As for every no-go result the key point are the assumptions so let us clearly state them:
\begin{enumerate}
\item First we assume that the D3 motion takes place along the angular direction at the tip of a deformed conifold. This is in contrast to the most studied case where the D3-brane moves along the radial direction \cite{Kachru:2003sx}. There are three important differences: the first is that the radial D3-brane coordinate is a conformally coupled scalar in the 4D effective action \cite{Kachru:2003sx,Buchel:2003qj} while the angular coordinates are not; this implies that the large mass, generated by the stabilization effects, for the radial D3-brane inflation is absent when the motion is along the base of the conifold. Whether the slow roll conditions are satisfied depends on the details of the potential \eqref{pot2}. As we discussed in section \ref{subsec:hilltop}, for the simplest embedding \eqref{kup} this can be achieved by fine tuning the stringy parameters. It seems plausible that more generic embeddings could present flat regions. We leave the investigation of this possibility for future research. The second difference is that the warp factor (and consequently the sound speed) varies in the radial direction but does not depend on the angular position. The parameter $f$ characterizing the DBI inflation is a constant at the tip and not a function of the inflaton. This feature was already considered in \cite{Kecskemeti:2006cg,Chen:2005fe} but was obtained there as an approximation of the radial motion of a D3-brane close to the tip of the deformed conifold where the AdS solution is not valid anymore. The third difference is that the field range for the angular motion is smaller than for the radial motion. In view of the discussions in \cite{Baumann:2006cd}, this implies that in these models, only negligible tensor perturbations can be produced during slow roll inflation.
\item \label{ass2} Second, we assume that since after the CMB perturbations are produced all the way until the end of inflation, the inflaton motion is exclusively relativistic, i.e. 
\be 
\frac{V^{'2}}{3V}f\Mp^2\gg 1 \qquad \Rightarrow \qquad \gamma\gg1 \nonumber\,.
\ee
This condition selects a class of potential or equivalently a class of D7 embeddings $g(z)$ leading to these potentials. As we show in the following, \textit{potentials in this class do not give enough e-foldings of inflation at the tip}. It is this assumption that we will relax in the next section where we will provide a simple example of a potential that possesses both DBI and slow roll regions. We will argue that these kind of potentials are promising candidates for a successful model of D-brane inflation.
\end{enumerate}
The argument is now based on the impossibility, under these two assumptions, to obtain enough e-foldings. Consider
\be \label{ne}
N_e&=&\int d\phi \frac{H}{\dot{\phi}}=\int \frac{d\phi}{\Mp}\sqrt{\frac{V^2}{V^{'2}}+\frac{fV\Mp^2}{3}}\nonumber\\
&\simeq&\int \frac{d\phi}{\Mp}\sqrt{\frac{fV}{3}}\lesssim \frac{\Delta \phi}{\Mp}\sqrt{\frac{fV_{max}}{3}}
\ee
where $\dot{\phi}$ and $H$ come from \eqref{phidot} and \eqref{H} respectively and we have used assumption \ref{ass2} to neglect the ``slow roll'' contribution $V^2/V^{'2}$ to $N_e$.
To trust our effective action description of string theory we have to require that the energy density during inflation is much smaller than the redshifted string scale: $V_{max}\ll h^4 M_s^4$. The parameter $f$ in \eqref{action} can be written in terms of stringy parameters as $h^{-4}T_3^{-1}$. Hence
\be 
\left\lbrace \begin{array}{c}
              V_{max}\ll h^4 M_s^4 \\
		f=\left(h^4T_3\right)^{-1}=\frac{(2\pi)^3 g_s}{h^4 M_s^4}
             \end{array}
\qquad \Rightarrow \qquad \sqrt{V_{max}f}\ll \sqrt{\frac{(2\pi)^3 g_s}{3}}\simeq9\sqrt{g_s} \right.\,.\nonumber
\ee
The field variation $\Delta\phi$ during inflation is smaller than $\pi\dd$. As we argued at the end of section \ref{subsec:slowroll}, generically $\dd\ll\Mp$ and therefore $\Delta \phi/\Mp<1$. Putting these bounds together we conclude that it \textit{is impossible to have more than a few e-foldings from exclusively DBI inflation at the tip}.

%%%%%%%%%%%%%%%%%%%%%%%%%%%%%%%%%%%%%%%%%%%%%%%%%%%%%%%%%%%%%%%%%%%%%%%%%%%%%%%%%%%%%%%%%%%%%%%%%%%%%%%%%%%%%%%

\subsection{Slow roll-DBI alternation from generic embeddings}\label{no go2}

A way to get around this negative result is to relax assumption \ref{ass2}, i.e. that during the whole 60 e-foldings of the observable inflation the motion is relativistic. Then in certain regions of the tip the potential can be of the slow roll type and the motion becomes non relativistic; there the ``slow roll'' term $V^2/V^{'2}$ can give a large contribution to $N_e$ in \eqref{ne} increasing the number of e-foldings.

For a generic Kuperstein embedding defined by an arbitrary holomorphic function $\g$, the potential was given in \eqref{gen}. The fact that Kuperstein embeddings preserve an $SO(2)$ symmetry implies that $V$ depends only on three real coordinates $\sigma,\,x_1$ and $\xx$ (up to $SO(4)$ permutations). The volume can be fixed \`{a} la KKLT and we have to study $V(\sigma_0(\phi,\xx),\phi,\xx)$. 

As a result of our discussion, we should look for functions $\g(\zz)$ such that the resulting scalar potential is of the slow roll type somewhere on the tip. This can give us enough e-foldings avoiding the no go theorem of the last section. Depending on the explicit form of $\g$, the perturbations could be produced during a DBI or a slow roll phase. We leave a deeper investigation of this promising model for the future. Here we comment on an alternative possibility.

Consider the embedding
\be\label{333}
g(z)=1-\frac{z_1^3}{\mu}\,,
\ee
where for simplicity we take $\mu$ to be real. In a non compact conifold, \eqref{333} defines three disconnected 4 cycles, each one described by the embedding $z_1=e^{i2\pi j/3}\mu^{1/3}$ for $j=0,1,2$ respectively. If a stack of D7-branes is wrapped around each 4-cycle the configuration is supersymmetric because all the three stacks respect the same supersymmetries as the background. We then suppose that it is possible to ``cut'' the conifold and ``glue'' it to a compact Calabi Yau; in addition, we ask that these three 4-cycles become part of a single 4-cycle. These are strong assumptions and we do not have anything to say about the hard problem of showing that the above construction can be actually realized. Here we would just like to present an interesting feature of this configuration. 

\begin{figure}
\centering
\includegraphics[width=1\textwidth,height=0.4\textwidth]{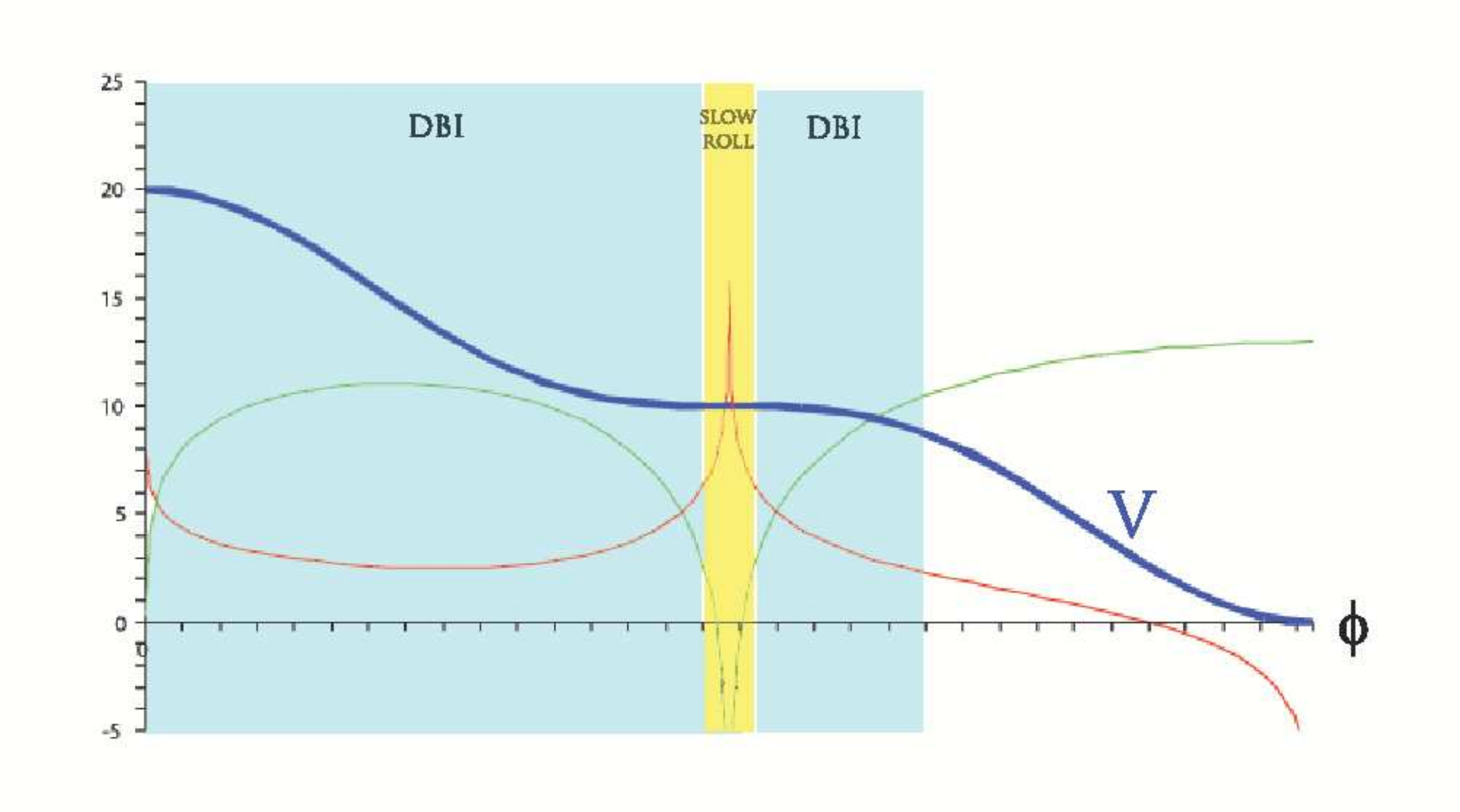}
\caption{The thick (blue) line is the potential in \eqref{3}. The two thin lines are the two DBI conditions \eqref{cond1} and \eqref{cond2} plotted in logarithmic scale. When they approach zero the DBI conditions are no more satisfied and the relativistic inflation regime end. Around the inflection point first and second derivative of the potential are very small and slow roll (non relativistic) inflation can take place. We highlight with different colors the DBI and slow roll phases.\label{fig:cos3}}
\end{figure}

So in the global compact picture, there is just one stack of D7-branes that wraps a single 4-cycle. Gaugino condensation on these D7's can give rise to a superpotential that, for a D3-brane inside the throat, is well approximated by
\be 
%W&=&W_0+A_0 \prod_{j=0}^2 \left(1-e^{i2\pi j/3}\frac{z_1^3}{\mu^{1/3}}\right)^{1/n} e^{-a\rho}\nonumber\\
W&=&W_0+A_0\left(1-\frac{z_1^3}{\mu}\right)^{1/n} e^{-a\rho}\,.
\ee
The same expansion discussed in section \ref{sec:kup} leads to an effective potential of the form
\be \label{3}
V\simeq\Lambda\left( 1+\cos^3\frac{\phi}{\dd}\right)\,,
\ee
that we plot in figure \eqref{fig:cos3}. The potential presents an inflection point where first and second derivatives vanish and slow roll inflation can take place. 

If the first derivative at the inflection point is strictly vanishing, then the inflaton reaches it only in an infinite time and inflation lasts forever. There are two possibilities to regularize this divergence: as discussed in \cite{Krause:2007jk}, the initial speed of the inflaton (or equivalently corrections to the strict slow roll regime) might allow it to pass the inflection point in a finite time. A second competing effect are the subleading terms that we neglected in the potential; they can have a non vanishing first derivative at the inflection point. If it has the right sign, the total number of e-foldings is finite, otherwise a local minimum is formed.

The CMB perturbations in this model can be produced both in the DBI or in the slow roll phase. In the first case, the analysis is very similar to the one we performed in section \ref{subsec:nut1}. The slow roll phase would then only provide enough e-foldings and would not have other observable consequences. In the second case, inflation takes place mainly in the slow roll regime, except for the last few unobservable e-foldings of DBI phase (on the right hand side of figure \eqref{fig:cos3}). The analysis is then very similar to other cases of inflection point inflation in the literature \cite{Krause:2007jk,Baumann:2007ah,Allahverdi:2006we,Panda:2007ie}. The phenomenological predictions are very sensitive to the parameters, e.g. to those determining the first derivative of the potential at the inflection point. 

Summarizing, \textit{inflation at the tip is a promising candidate for a phenomenologically successful embedding of DBI inflation in string theory}. The right perturbations can be easily produced; to obtain enough e-foldings it is required that, in some region, the potential is slow roll flat. We have given above a (semi) explicit example, but we expect that many others can be found studying more generic D7's embeddings (e.g. with a generic $\g$ in \eqref{kup}).

%%%%%%%%%%%%%%%%%%%%%%%%%%%%%%%%%%%%%%%%%%%%%%%%%%%%%%%%%%%%%%%%%%%%%%%%%%%%%%%%%%%%%%%%%%%%%%%%%%%%%%%%%%%%%%%
% \section{Initial condition, graceful exit and reheating}\label{reheating}
% 
% A last comment on graceful exit and reheating is in order. Differently from the D3 anti D3 inflation model, here there is no annihilation event that stops inflation. For inflation at the tip there are two ways in which graceful exit can take place: when the inflaton reaches the end of the potential, the slow roll conditions are not anymore fulfilled; the oscillations of the inflaton condense and subsequently decay as in the standard picture. Alternatively, if more than one brane is present at the tip, as might be natural as a result of brane flux annihilation \cite{DeWolfe:2004qx}, assisted inflation will take place. All the D3-branes are driven to the same minimum where they will collide\footnote{For the simplest Kuperstein embedding, $\g=\mu$, there is only one (supersymmetric) minimum. For a generic $\g$ this is not the case anymore and there can be a family of vacua. A classification of supersymmetric vacua for various $\g$ was given in \cite{DeWolfe:2007hd}.}. The study of \cite{Kofman:2004yc} shows that as result of the collision, the kinetic energy of the branes is transferred to the gauge theory living on the newly formed stack of branes. This D-brane trapping provides a graceful exit and suggests a mechanism for reheating the world volume theory living on the brane.

%%%%%%%%%%%%%%%%%%%%%%%%%%%%%%%%%%%%%%%%%%%%%%%%%%%%%%%%%%%%%%%%%%%%%%%%%%%%%%%%%%%%%%%%%%%%%%%%%%%%%%%%%%%%%%%

\section{Conclusions and perspectives}\label{sec:con}

In this paper we show how to construct a model of D3-brane inflation that takes place along the angular directions of a warped conifold and where no anti D3-branes are needed. A D3-brane moves on the tip of a warped deformed conifold \cite{Klebanov:2000hb} embedded in a compact Calabi Yau, which is an $S^3$ (see figure \ref{cartoon}). The potential comes from the F-term where the threshold corrections to the non perturbative superpotential are taken into account.

The potential is determined by the supersymmetric embedding of the stack of D7-branes generating the non perturbative superpotential. We write down the scalar potential for a class of supersymmetric embeddings \cite{Kuperstein:2004hy} defined by an arbitrary holomorphic function $\g(z)$. We concentrate on the simplest which is a constant $\g(z)$. The potential (for the canonically normalized inflaton) takes then the form
\be\label{nat inf end}
V(\phi)= \Lambda \left(1+ \cos\frac{\phi}{\dd} \right)+C\sin^2\frac{\phi}{\dd}\,.
\ee
We have looked for inflationary trajectories, considering two classes:
\begin{enumerate}
\item \textbf{Slow roll inflation}: it is possible to achieve but only allowing for fine tuning. In a large region of the parameter space, $C$ in \eqref{nat inf end} is much smaller than $\Lambda$, so that $V(\phi)$ takes the form of the Natural Inflation potential \eqref{nat infl pot} \cite{Savage:2006tr}. Other models tried to embed Natural Inflation in string theory proposing some axions as the candidate inflaton \cite{Grimm:2007hs}. Difficulties arise because the data require \cite{Savage:2006tr} a large axion decay constant (that we indicate with $\dd$, see \eqref{nat infl}), which has not been found in a controllable string theory model. In our model we encounter the same problem, generically $\dd<\Mp$.
On the other hand, in a small region of the stringy parameter space, we can have $\Lambda\simeq 2C$ and the potential becomes very flat on the top (see figure \ref{fig2}). This is a kind of hilltop model \cite{Kohri:2007gq}, i.e. of the form 
\be
V\sim \Lambda-\frac12 m^2\phi^2-\lambda \phi^4\,.
\ee
The negative square mass of the inflation can be made small by \textit{fine tuning}, and we obtain a \textit{string theory model of slow roll inflation in perfect agreement with CMB data} (see section \ref{subsec:hilltop}). The model predicts $n_s\simeq0.94$, negligible tensor modes and the scale of inflation $(\Lambda)^{1/4}\sim\dd \times10^{-3}$. 

\item \textbf{DBI inflation}: as the inflaton corresponds to the position of a D3-brane, the kinetic term comes from the DBI action. This can allow inflation to take place even with a steep potential \cite{Alishahiha:2004eh}. We investigate this possibility and discover that, although \textit{perturbations compatible with CMB data can be produced}, \textit{the DBI regime can never lead to more than a few e-foldings}. We argue that more general embeddings than the one we studied here (Kuperstein embedding \cite{Kuperstein:2004hy} with constang $\g$) can possess flat regions that would increase the number of e-foldings and lead to a successful model. We leave a more detailed study of this interesting possibility for future investigation.
\end{enumerate}

There other two results that we would like to emphasize, which are somehow outside the main focus of the paper. The first is that in the presence of the DBI kinetic term, the Natural Inflation potential can lead to a phenomenologically viable inflation also for \textit{subplanckian values of the axion decay constant} $\dd$ (in fact as low as $0.04\Mp$).

The second regards radial brane-anti brane inflation. We propose a way to cancel the inflaton mass, induced by moduli stabilization effects, using the uplifting term. Certain conditions on the volume scaling of the uplifting are required for this mechanism to work. We refer the reader to appendix \ref{app:up} for more details on these \textit{``nice'' upliftings}.

In the following we list some interesting directions for future research
\begin{itemize}
\item The case of a generic embedding (with a generic $\g$ in \eqref{kup} or those discussed in \cite{Ouyang:2004}) has to be studied in detail looking for an explicit working model. A criteria to keep in mind in this search follows from our no go result in section \eqref{no go}: a successful model needs to have a potential with at least one slow roll flat region.
\item The absence of an anti D3-brane makes the mechanism of graceful exit and reheating in the present model very different from the standard D3 anti D3 inflation model. A mechanism naturally embedded in the model is the D-brane trapping (which was one of the motivations of the original proposal of brane inflation \cite{Dvali:1998pa}). In \cite{Kofman:2004yc} it was shown that in the collision of branes, the kinetic energy can be transferred to the gauge fields living on the newly created stack. This would create a thermal bath in the world volume of the branes that could evolve into our universe. It would be interesting to investigate quantitatively the phenomenological viability of this idea. 
\item It would be desirable to develop a gauge dual description of the brane inflation at the tip proposed in this paper, a work in this direction is \cite{Dymarsky:2005xt} (see also \cite{Buchel:2006em}). Although there the origin of the potential is different, the authors provide a gauge theory description of the radial motion of a brane in a resolved warped deformed conifold. 
\item In appendix \ref{app:up} we propose a new mechanism to fine tune the inflaton mass in radial brane inflation using the uplifting term. It would be interesting to construct an explicit string model that produces such ``nice'' uplifting and thoroughly study its phenomenological implications for inflation (e.g. the effects of the quartic term in \eqref{nice}, etc\dots).
\end{itemize}

%%%%%%%%%%%%%%%%%%%%%%%%%%%%%%%%%%%%%%%%%%%%%%%%%%%%%%%%%%%%%%%%%%%%%%%%%%%%%%%%%%%%%%%%%%%%%%%%%%%%%%%%%%%%%%%

\section*{Acknowledgements}
%\ack

It is a pleasure to thank Thomas Grimm, Michael Haack, Renata Kallosh, Luca Martucci and Liam McAllister for helpful discussions. This work is supported by the DFG within the Emmy Noether-Program (grant number: HA 3448/3-1).

%%%%%%%%%%%%%%%%%%%%%%%%%%%%%%%%%%%%%%%%%%%%%%%%%%%%%%%%%%%%%%%%%%%%%%%%%%%%%%%%%%%%%%%%%%%%%%%%%%%%%%%%%%%%%%%

\appendix

\section{The deformed conifold}\label{app:def}

In this appendix we review some useful facts about the geometry of the deformed conifold. We follow \cite{Candelas:1989js,Minasian:1999tt}. At the end we will obtain an expression for $\partial_r z_1$ at the tip, which we will use to prove the radial stability in \ref{app:radial}.

The deformed conifold is defined by the following hypersurface in $\mathbb{C}^4$
\be\label{coni2}
\sum_{A=1}^4(z_A)^2=\varepsilon^2\,,
\ee 
where $\varepsilon$ is a complex parameter. In terms of the matrix $W=\frac{1}{\sqrt{2}} (z_i \sigma_i+iz_4)$, this expression can be written as
\be 
\mathrm{det}W=-\frac{\varepsilon^2}{2}
\ee
A radial coordinate can be defined by
\be
\sum_{A=1}^4|z_A|^2=\mathrm{tr}\left(WW^{\dagger}\right)= r^3\,.
\ee
A generic solution to \eqref{coni2} can be written as 
\be 
W=L_1Z_0L_2^{\dagger}\,,
\ee
where $L_1\,, L_2$ are $SU(2)$ matrices that can be parameterized using Euler angles
\be 
L_j=\left( 
	\begin{array}{cc}
		\cos\frac{\theta_j }{ 2} e^{i(\psi_j+\phi_j)/2} & -\sin\frac{\theta_j }{2}e^{-i(\psi_j-\phi_j)/2} \\
		\sin\frac{\theta_j }{ 2} e^{i(\psi_j-\phi_j)/2} &  \cos\frac{\theta_j }{2}e^{-i(\psi_j+\phi_j)/2} 
	\end{array}
\right)
\ee 
and 
\be 
Z_0&=&\left( 
	\begin{array}{cc}
	 	0 & a \\
		b & 0 
	\end{array}
\right)\,,\\
a&=&\frac12\left(\sqrt{r^3+\varepsilon^2}+\sqrt{r^3-\varepsilon^2}\right)\,,\qquad b=\frac{\varepsilon^2}{2a}\,.
\ee
For $r^{3/2}>\varepsilon$ one of the six angles $\{\phi_i,\psi_i,\theta_i\}$, $i=1,2$, is redundant. We can fix the gauge imposing $\psi_2=0$; we obtain the following real parameterization of the complex coordinates $z_A$
\be
z_1&=& \frac{1}{\sqrt {2}} \left( a\cos \frac{\theta_1}{2} \cos \frac{\theta_2}{2} e^{i\left( \psi_1+\phi_1+\phi_2 \right)/2}-b\sin \frac{\theta_1}{2} \sin \frac{\theta_2}{2} e^{-i \left( \psi_1-\phi_1-\phi_2 \right)/2 } \nonumber \right.  \\
& & \left. - a\sin \frac{\theta_1}{2} \,\sin \frac{\theta_2}{2} e^{i \left( \psi_1-\phi_1-\phi_2 \right)/2 }+b\cos \frac{\theta_1}{2} \cos \frac{\theta_2}{2} e^{-i \left( \psi_1+\phi_1+\phi_2 \right)/2 } \right) \nonumber \,,\\
z_2&=& \frac{i}{\sqrt {2}} \left( a\cos \frac{\theta_1}{2} \cos \frac{\theta_2}{2} e^{i\left( \psi_1+\phi_1+\phi_2 \right)/2}-b\sin \frac{\theta_1}{2} \sin \frac{\theta_2}{2} e^{-i \left( \psi_1-\phi_1-\phi_2 \right)/2 } \nonumber \right. \\
& & \left. + a\sin \frac{\theta_1}{2} \,\sin \frac{\theta_2}{2} e^{i \left( \psi_1-\phi_1-\phi_2 \right)/2 }-b\cos \frac{\theta_1}{2} \cos \frac{\theta_2}{2} e^{-i \left( \psi_1+\phi_1+\phi_2 \right)/2 } \right) \nonumber\,,\\
z_3&=& -\frac{1}{\sqrt {2}} \left( a\cos \frac{\theta_1}{2} \sin \frac{\theta_2}{2} e^{i\left( \psi_1+\phi_1-\phi_2 \right)/2}+b\sin \frac{\theta_1}{2} \cos \frac{\theta_2}{2} e^{-i \left( \psi_1-\phi_1+\phi_2 \right)/2 } \nonumber \right. \\
& & \left. - a\sin \frac{\theta_1}{2} \,\cos \frac{\theta_2}{2} e^{i \left( \psi_1-\phi_1+\phi_2 \right)/2 }-b\cos \frac{\theta_1}{2} \sin \frac{\theta_2}{2} e^{-i \left( \psi_1+\phi_1-\phi_2 \right)/2 } \right) \nonumber \,,\\
z_4&=& \frac{i}{\sqrt {2}} \left( -a\sin \frac{\theta_1}{2} \cos \frac{\theta_2}{2} e^{i\left( \psi_1-\phi_1+\phi_2 \right)/2}-b\cos \frac{\theta_1}{2} \sin \frac{\theta_2}{2} e^{-i \left( \psi_1+\phi_1-\phi_2 \right)/2 } \nonumber \right. \\
& & \left.  a\cos \frac{\theta_1}{2} \,\sin \frac{\theta_2}{2} e^{i \left( \psi_1+\phi_1-\phi_2 \right)/2 }+b\sin \frac{\theta_1}{2} \cos \frac{\theta_2}{2} e^{-i \left( \psi_1-\phi_1+\phi_2 \right)/2 } \right) \nonumber \,.\\
\ee
The singular conifold is just the special case $\varepsilon=0$ that implies $a=r^{3/2}\,,b=0$. Now we want to calculate the derivative of $z_1$ with respect to $r$ evaluated at the tip $r=\varepsilon^{2/3}$. At the tip three of the six Euler angles are redundant, a possible choice of the gauge is $\theta_2=\psi_2=\phi_2=0$. The result is then
\be 
\lim_{r\rightarrow\varepsilon^{2/3}}\mathrm{Im} \frac{\partial z_1}{\partial r}&=& \frac{\sqrt{6}}{4}\frac{\cos{\frac{\theta_1}{2}} \sin \frac{\psi_1+\phi_1}{2} \varepsilon^{2/3} }{\sqrt{r-\varepsilon^{2/3}}}+\mathcal{O}(1) \,, \nonumber \\
\lim_{r\rightarrow\varepsilon^{2/3}}\mathrm{Re} \frac{\partial z_1}{\partial r}&=& \frac{3}{4}\varepsilon^{1/3}\cos \frac{\theta_1}{2}\cos\frac{\psi_1+\phi_1}{2}+\mathcal{O}\left( r-\varepsilon^{2/3} \right) \label{diffz} \\
\nonumber &=& \frac{3}{4}\varepsilon^{1/3} E(\theta_1,\phi_1,\psi_1)+\mathcal{O}\left( r-\varepsilon^{2/3} \right)\,,
\ee
where in the last line we have introduced the function $|E(\theta_1,\phi_1,\psi_1)|<1$ for further reference. We see that the imaginary part of the derivative is singular but the real part has a finite non vanishing value.

%%%%%%%%%%%%%%%%%%%%%%%%%%%%%%%%%%%%%%%%%%%%%%%%%%%%%%%%%%%%%%%%%%%%%%%%%%%%%%%%%%%%%%%%%%%%%%%%%%%%%%%%%%%%%%%

\section{Minimization of $\sigma$}\label{min}

% \begin{figure}
% \includegraphics[width=0.8\textwidth]{psi-sigma}
% %\includegraphics[height=6cm,width=0.9\textwidth]{psi-sigma2.eps}
% \includegraphics[width=0.8\textwidth]{psi-sigma3}
% \caption{We plot the potential as a function of $\sigma$ and $\psi$. This confirms graphically that the minimum in the $\sigma$ direction is almost independent of the value of $\psi$. Also the mass of the modulus $\sigma$ is much larger than that of $\psi$.}
% \end{figure}

The KKLT minimum of $\sigma$ \cite{Kachru:2003aw} is changed both by the uplifting and by the introduction of the D3-brane. Here we provide an analytical estimate along the lines of \cite{Krause:2007jk}. We call $\sigma_0$ the minimum of the KKLT potential with
\be\label{W0}
W_0=-A_0e^{-a\sigma_0}\left[1+\frac13 a (2\sigma_0-k_0) \right]\,,
\ee
$\sigma_{cr}$ is the actual minimum and $\Delta$ is the shift due to the uplifting and the D3-brane: $\Delta=\sigma_{cr}-\sigma_0$. We parameterize the uplifting as
\be \label{D}
V_{up}=\frac{D}{U^\bb}=\beta\frac{|A_0|^2a^2e^{-2a\sigma_0} (2\sigma_0-k_0)^{b-1}}{3U^b}\,,
\ee
where $\beta \gtrsim 1$ gives a de Sitter space. We substitute these expressions for $W_0$ and $D$ into the potential \eqref{pot2} and take the derivative with respect to $\sigma$. We substitute $\sigma=\sigma_0+\Delta$ in $V'=0$ and solve for $\Delta$. At leading order the result is given by
\be \label{delta}
\Delta=\frac{b}{2}\frac{\beta}{a^2\sigma_0}-\frac{\varepsilon }{a \mu n}\cos \phi+\frac{\varepsilon^{2/3}}{2c a^3 n^2\mu^2\sigma_0\gamma}\sin^2\phi
\ee
In the particular case $b=2$, the shift due to the uplifting reproduces the result found in \cite{Krause:2007jk,Baumann:2007ah}; we define $\sd\equiv\sigma_0+b\beta/(2a^2\sigma_0)$ that would be the minimum of the volume in the presence of an uplifting term but without any D3-brane. The first of the two $\phi$ dependent terms in \eqref{delta} is analogous to the dependence on the radial motion found in \cite{Krause:2007jk,Baumann:2007ah}. Finally the critical value of the K\"ahler modulus is
\be\label{scr}
\sigma_{cr}(\phi)\simeq \sd - \frac{\varepsilon}{a \mu n}\cos \phi+\frac{\varepsilon^{2/3}}{2c a^3 n^2\mu^2\sigma_0\gamma}\sin^2\phi \,.
\ee
We use this result to prove that the dependence of $\sigma_{cr}$ on $\phi$ is mild in the sense that it produces in the effective potential $V(\sigma_{cr}(\phi),\phi)$ only subleading terms. We substitute $W_0$ from \eqref{W0}, $D$ from \eqref{D} and $\sigma_{cr}$ from \eqref{scr} into $V(\sigma_{cr}(\phi),\phi)-V(\sd,\phi)$. We keep the leading terms in the $\varepsilon/\mu$ and large volume expansion, the result is
\be
V(\sigma_{cr}(\phi),\phi)-V(\sd,\phi)&\simeq&\frac{|A_0|^2 \varepsilon^{5/3}e^{-2a\sigma_0} }{6cn^3\mu^3 \sigma_0^2 \gamma}\cos \phi \sin^2 \phi+ \nonumber\\
&& \hspace{1cm} -\frac{|A_0|^2 a^2 \varepsilon^2 e^{-2a\sigma_0}}{6n^2\mu^2 \sigma_0}\cos^2\phi+\dots\,,\nonumber
\ee
Comparing these term with those in \eqref{L},\eqref{B} and \eqref{C}, we see that they are subleading in $\varepsilon/\mu$ or in the large volume expansion. We conclude therefore that at leading order $ V(\sigma_{cr}(\phi),\phi)\simeq V(\sigma_{up},\phi)$, i.e. we are allowed to neglect the $\phi$ dependence of $\sigma_{cr}$. This justify the use of \eqref{ex} or \eqref{33} as leading order potential.

%%%%%%%%%%%%%%%%%%%%%%%%%%%%%%%%%%%%%%%%%%%%%%%%%%%%%%%%%%%%%%%%%%%%%%%%%%%%%%%%%%%%%%%%%%%%%%%%%%%%%%%%%%%%%%%

\section{Radial stability}\label{app:radial}

In this appendix we analyze the radial stability for a D3-brane at the tip of a deformed conifold. We want to find the condition to satisfy
\be\label{r stab}
\frac{\partial}{\partial r} V(z)\mid_{r=\varepsilon^{2/3}}>0
\ee
The potential \eqref{pot}, as we already discussed, is valid only at the tip. There actually are some other terms, coming from the off diagonal elements of the K\"ahler metric \eqref{Kji} or the additional term in $K^{\bar{1}1}$ in \eqref{Kji} that are zero at the tip, but whose $r$ derivative might be non vanishing. The additional term in $K^{\bar{1}1}$ has two factors, $k_{\bar l}$ and $k_{h}$, that vanish at the tip, therefore also the $r$ derivative vanishes there. Only the off diagonal elements of the K\"ahler metric \eqref{Kji} contributes and we call these additional terms $V_{off}$ for ``off diagonal''. 

The scalar potential depends on the radial position of the D3-brane $r$ via the K\"ahler potential in \eqref{K} and the non perturbative superpotential $W_{np}$. We decompose $\partial_r V$ in four terms, the first three coming from \eqref{pot} and the last one from $V_{off}$:
\begin{enumerate}
\item All terms in \eqref{pot} coming from the $r$ dependence in $U(r)$. Using the explicit metric of the deformed conifold near the tip we have
\be
U&=&2\sigma-\gamma k(z,\bar z) \\
&\nonumber \simeq &2\sigma-\gamma \frac{c}{\varepsilon^{2/3}} \left(k_0+ \sum_{A=1}^{4} |z_A|^2 \right)= 2\sigma-\gamma \frac{c}{\varepsilon^{2/3}} \left(k_0+ r^3\right)\,.
\ee
We will indicate this contribution as
\be 
V_{r1}\equiv \frac{\partial U}{\partial r}\frac{\partial}{\partial U} \left( V_{up}+V_{KKLT}+\Delta V \right)\,.
\ee
\item All terms in \eqref{pot} coming from $\partial_r \Delta V$ where we neglect the terms coming form $\partial_r U(r)$ that appear already in $V_{r1}$. We will indicate this contribution as
\be 
V_{r2}\equiv\frac{\partial}{\partial r}\Delta V (U=2\sigma-\gamma k_0)\,.
\ee
\item All the remaining terms from \eqref{pot}, i.e. those coming from $\partial_{r} A(z)$ in $V_{KKLT}$. We will indicate this contribution as
\be
V_{r3}\equiv \frac{\partial A(z)}{\partial r}  \frac{\partial }{\partial A(z)} V_{KKLT}\,.
\ee
\item Finally the term $\partial_r V_{off}$ evaluated at the tip. It comes from the off diagonal terms
\be 
e^{\kq K}\left(K^{\bar \rho i} \overline{D_{\rho}W } D_iW + c.c.\right)\,.
\ee
To get a non vanishing contribution the $r$ derivative has to hit the $k_{l}$ in $K^{\bar \rho i} $, therefore
\be 
V_{r4}\equiv 2 \frac{\kq }{3U^2} W_i \overline{D_{\rho}W} k^{\bar{l}i}\frac{\partial}{\partial r} k_{\bar {l}}\,.
\ee
\end{enumerate}

Let us start with $V_{r1}$. Using \eqref{pot2},\eqref{pot3} and \eqref{Vup} it is straightforward to obtain a long expression for $V_{r1}$. To estimate it, we use the parameterization \eqref{D} and \eqref{W0} for $D$ and $W_0$. Then we substitute the minimum for the volume estimated in \eqref{scr}, i.e. $\sigma\simeq\sd$. Finally we look at the leading terms in the $\varepsilon/\mu$ and large volume expansion; the result is
\be\label{V1}
V_{r1}\simeq\frac{\kq |A_0|^2a^2\gamma c e^{-2a\sigma_0}\varepsilon^{2/3}}{4\sigma_0^2}\left[(b\beta-3)+\frac{b\beta}{a\sigma_0}+\frac{\varepsilon^{4/3}}{cn^2\mu^2a^2\sigma_0}+\dots\right]\,.
\ee
To calculate $V_{r2}$, we substitute in \eqref{pot3} the embedding \eqref{np kup} and $U=2\sigma-\gamma k_0$. Then using the chain rule (and the fact that for constant $U$, $\partial_{x_2}\Delta V=\partial_{x_3}\Delta V=\partial_{x_4}\Delta V=0$) we can write
\be 
V_{r2}=\frac{\partial x_1}{\partial r} \frac{\partial }{\partial x_1} \Delta V \simeq \frac34 \varepsilon^{1/3} E \frac{\partial }{\partial x_1} \Delta V\,,
\ee 
where we used the result \eqref{diffz} of appendix \ref{app:def}; remember that $-1<E(\phi_1,\theta_1,\psi)<1$. Following the same steps as for $V_{r1}$, we can estimate $\Delta V_{x_1}$; the result is
\be \label{V2}
V_{r2}\simeq -\frac{|A_0|^2e^{-2a\sigma_0} E \cos \phi}{8\mu^2n^2\gamma c \sigma_0^2}+\dots
\ee
For $V_{r3}$ we obtain
\be \label{V3}
V_{r3}&=&\frac{\partial x_1}{\partial r}\frac{\partial A(x_1)}{\partial x_1}\frac{\partial}{\partial A(x_1)}V_{KKLT}\nonumber\\
&=&\frac{|A|^2ae^{-2a\sigma_0}\varepsilon^{1/3}}{8\mu n \sigma_0^2} E \left[ (b\beta-3)+\frac{b\beta(14-3b\beta)}{4a\sigma_0}+\dots \right]\,.
\ee
Finally we come to $V_{r4}$. Using again the chain rule (it is not necessary to distinguish $\partial_{z_i}$ from $\partial_{\bar z_i}$ because they appear symmetrically in $k(z,\bar z)$ and at the tip they are both equal to $x_i$) we obtain
\be 
V_{r4}&=& 2\frac{\kq \overline{D_{\rho}W}}{3U^2} W_i  k^{\bar{l}i}\frac{\partial z_j}{\partial r} \frac{\partial}{\partial z_j}k_{\bar {l}}\,=2\frac{\kq \overline{D_{\rho}W}}{3U^2}\frac{\partial z_j}{\partial r} \delta_{ij}W_j \nonumber \\
&=&2\frac{\kq \overline{D_{\rho}W}}{3U^2}\frac{\partial z_1}{\partial r} \frac{A(x_1)e^{-a\sigma}}{n(\mu-x_1)} \nonumber \\
&\simeq&\frac2 34 E\varepsilon^{1/3}\frac{A_0e^{-a\sigma_0}}{n\mu}\frac{\beta b A_0 e^{-a\sigma_0}}{3(2\sigma_0)^3}\nonumber = \frac{|A|^2 a e^{-a\sigma_0}\varepsilon^{1/3}}{8\mu n \sigma_0^2}\, E \, \frac{b\beta }{2a\sigma_0}\,,
\ee
from which it is transparent that $V_{r4}$ is of the same order as the subleading term of $V_{r3}$ in \eqref{V3}.

We want now to compare these four contributions to understand which is the leading one and in which regime. An interesting case to consider is $b=3$, because explicit uplifting mechanisms exist with this scaling \cite{Burgess:2003ic,Saltman:2004sn}. We are not aware of any explicit model with $b>3$ but it would be intersting to investigate this possibility, e.g. in the framework of F-term uplifting \cite{Brax:2007xq}. Other consideration about the role of the uplifting term can be found in appendix \ref{app:up}. 
As we argued in the main text, we will be intersted in $\beta\simeq1$ such that the cosmological constant after inflation is negligibly small. Then a subtlety arises from the fact that for $b=3$ and $\beta\simeq1$, the leading term of both $V_{r1}$ and $V_{r3}$ receives and additional suppression by the coefficient $(b\beta-3)$, which is not taken into account by the $\varepsilon/\mu$ and large volume expansion. We need therefore an estimate for $(\beta-1)$.

%%%%%%%%%%%%%%%%%%%%%%%%%%%%%%%%%%%%%%%%%%%%%%%%%%%%%%%%%%%%%%%%%%%%%%%%%%%%%%%%%%%%%%%%%%%%%%%%%%%%%%%%%%%%%%%

\subsection{Radial stability in the $C\ll B$ regime (DBI inflation)}

Consider the expression for $\Lambda$ in \eqref{L}. If we want the cosmological constant to be negligibly small after inflation, then we need $(\beta-1)$ to be of order $(a\sigma_0)^{-2}$, so that the leading term cancel the next to leading one. This tells us that in the case $b=3$, the $(b\beta-3)$ term in $V_{r1}$ and $V_{r3}$ becomes subleading with respect to the next to leading one which is suppressed by just a relative $(a\sigma_0)^{-1}$ factor.
We are now ready to compare the three contributions:
\be\label{radstab}
V_{r1} \gg V_{r3},V_{r4} & \Longleftrightarrow & (2\pi\gamma c\mu)\varepsilon^{1/3}\gg 1 \nonumber\\
V_{r1} \gg V_{r2} & \Longleftrightarrow & \left\lbrace 
					\begin{array}{lcl} 
					(2\pi\gamma c \mu)\varepsilon^{1/3}\gg 1 &\textrm{for}& \quad b\neq3 \\
					(2\pi\gamma c \mu)\varepsilon^{1/3}\gg \sqrt{a\sigma_0}&  \textrm{for}& \quad b=3
					\end{array} \right. \nonumber
\ee
where we have substituted $an=2\pi$. 

We remind the reader that the conditions which was obtained in section \ref{sec:simpl}
\be 
C\ll B & \Longleftrightarrow & \left\lbrace 
					\begin{array}{lcl} 
					(2\pi\gamma c \mu)\varepsilon^{1/3}\gg 1 &\textrm{for}& \quad b\neq3 \\
					(2\pi\gamma c \mu)\varepsilon^{1/3}\gg \frac{4}{15} a\sigma_0 &  \textrm{for} & \quad b=3
					\end{array} \right. \nonumber
\ee
It follows that in the regime $C\ll B$, which is the one we consider in section \ref{subsec:slowroll} and \ref{subsec:nut1}, $V_{r1}$ is always the leading contribution. From \eqref{V1} we see that the derivative in the radial direction is positive, i.e. a D3-brane at the tip will not move in the radial direction, as long as $b\geq 3$.

%%%%%%%%%%%%%%%%%%%%%%%%%%%%%%%%%%%%%%%%%%%%%%%%%%%%%%%%%%%%%%%%%%%%%%%%%%%%%%%%%%%%%%%%%%%%%%%%%%%%%%%%%%%%%%%

\subsection{Radial stability in the 2C=B regime (slow roll hilltop inflation)}

The regime $C=B/2$, that as we saw in section \ref{subsec:hilltop} can lead to slow roll hilltop inflation, requires a separate discussion. 

Let us start with the case $b=3$. As discussed in section \ref{subsec:slowroll}, to allow for slow roll inflation at the top of the potential we need to fine tune the parameters such that $2C=B=\Lambda$, which from \eqref{V2} and \eqref{V3} implies
\be 
2C=B \qquad \Longleftrightarrow \qquad (2\pi\mu c\gamma)\varepsilon^{1/3}=\frac{4}{15}a\sigma_0\,.
\ee
Using this new constraint the condition \eqref{radstab} for the radial stability becomes:
\be
0&<&V_{r1}+V_{r2}+ V_{r3}+V_{r4}=\frac{|A|^2e^{-2a\sigma_0}}{40\mu^2cn^2\gamma\sigma_0^2 }\left[\frac{32a\sigma_0}{15}+E(7-5\cos\phi)\right]\,.
\ee
Using the fact that $|E|<1$, one obtains that the above inequality is satisfied for $a\sigma_0>5.6$ and an arbitrary value of the angular coordinates. We remind the reader that if we want to be able to neglect higher instanton contributions to the nonperturbative superpotential we have to require $a\sigma_0\gg1$. Therefore, we conclude that in the case $2C=B$ and $b=3$ the D3-brane at the tip is (meta)stable in the radial direction.

Let us now consider the case $b>3$, then 
\be 
B=2C \qquad \Longleftrightarrow \qquad (2\pi c \mu\gamma)\varepsilon^{1/3}=1\,.
\ee
Using this constraint and taking for concreteness $b=4$, the condition \eqref{radstab} for the radial stability becomes:
\be 
0&<&V_{r1}+V_{r2}+ V_{r3}+V_{r4}=\frac{1}{8\mu^2cn^2\gamma\sigma_0^2 }\left[2+E(1-\cos\phi)\right]\,.
\ee
The inequality is fulfilled by any value of the angular coordinates, except for the case $E=-1$ and $\phi=\pi$ which saturates it. Anyway in the case we are considering $B=2C$, (slow roll) inflation takes place next to the top of the potential, where $(1-\cos\phi)$ is very small and therefore the inequality is confortably satisfied.

%%%%%%%%%%%%%%%%%%%%%%%%%%%%%%%%%%%%%%%%%%%%%%%%%%%%%%%%%%%%%%%%%%%%%%%%%%%%%%%%%%%%%%%%%%%%%%%%%%%%%%%%%%%%

\section{Nice upliftings}\label{app:up}

The setup of this appendix is different from the one in the rest of the paper. Also we will use a simplified notation which makes it easier to capture the features of the mechanism we are proposing. We consider a generalization of the KKLMMT model \cite{Kachru:2003sx} where the uplifting has an a priori generic scaling. We show that for a certain class of uplifting it is possible, allowing for fine tuning, to cancel the large mass of the inflaton that in the KKLMMT scenario induces a large $\eta$ and prevents slow roll inflation. 

Consider the potential 
\be 
V(\phi)=-\frac{|V_{KKLT}|}{U^2}+\frac{\tilde{D}}{U^b}\,,
\ee
where the uplifting has a general volume (and therefore inflaton) dependence parameterized by $b$, $V_{KKLT}$ is the KKLT potential \cite{Kachru:2003aw} and\footnote{Note that this definition is different from the one we use in the main text. The volume modulus $\sigma$ has already been stabilized. Here the quantities $V_{KKLT}$, $D$ and $U$ differ from the one in the main text just by some factor that is unimportant for the discussion in this appendix.} $U=1-\phi^2/(6\Mp^2)$. Up to an overall factor
\be 
V(\phi)=-\frac{1}{U^2}+\frac{D}{U^\alpha}\,,
\ee
therefore expanding for small $\phi$
\be \label{nice}
V(\phi)\simeq -1+D+\left( -\frac13+\frac16 D\alpha\right)\frac{\phi^2}{\Mp^2}+\dots\,.
\ee
It is clear that the mass term for the inflaton can be made vanishing with an appropriate choice of the uplifting. The requirement is $Db=2$. The condition for a de Sitter vacuum is $D \gtrsim 1$ (up to terms of order $\phi^2$). An uplifting with $b < 2$ allows at the same time to cancel the inflaton mass and uplift the AdS vacuum to a de Sitter one. In this minimal setup it is amusing to see that the cosmological is related to the mass of inflaton. This reduces by one with respect to KKLMMT (see e.g. \cite{Bean:2007hc}) the string parameters that we need to fix by experiment. If, for example $b$ is very close to $2$ but strictly smaller, then a small cosmological constant is equivalent to a small inflaton mass. 

Unfortunately there is no well understood uplifting mechanism with $b<2$. An anti D3-brane gives an uplifting that scales as $U^{-3}$. This generalizes for a Dp-brane wrapping a $p-3$ cycle to $U^{\frac{p-15}{4}}$ \cite{private}. If the cycle wrapped by the Dp-brane is at the tip of a warped throat, then the warp factor gives an additional $U$ factor. In KKLT for example, the anti D3-brane produces an uplifting term $U^{-2}$. It would be intersting to study the configuration in which, e.g. a D5-brane wraps a nonvanishing two cycle at the tip of the throat\footnote{it is not necessary that it is a topologically non trivial cycle but it could also be a metastable configuration, e.g. \cite{DeWolfe:2004qx}} leading to $b=3/2$.

%%%%%%%%%%%%%%%%%%%%%%%%%%%%%%%%%%%%%%%%%%%%%%%%%%%%%%%%%%%%%%%%%%%%%%%%%%%%%%%%%%%%%%%%%%%%%%%%%%%%%%%%%%%

\subsection{Nice downliftings}

Another possibility of \textit{nice upliftings} is when there are at least two uplifting terms of opposite sign\footnote{A more evocative name for the negative one would be ``downlifting''.}. Up to some factors, the potential is of the form
\be 
V(\phi)&=&-\frac{1}{U^2}+\frac{D_1}{U^{b_1}}\,-\frac{D_2}{U^{c_2}}\\
&=&D_1-D_2-1+\left( -\frac13+\frac16 D_1b_1-\frac16 D_2b_2\right)\phi^2+\dots\,.
\ee
where $D_1,D_2>0$. Then, to cancel the inflaton mass one has to require
\be 
m^2_{\phi}=0\qquad&\Rightarrow&D_1=\frac{2+D_2b_2}{b_1} \nonumber\\
V>0\qquad&\Rightarrow&D_1>D_2+1\,.
\ee
A simple example is $b_1=2,b_2=3$, then fine tuning $D_1=1+D_23/2$ cancels the inflaton mass. It would be intersting to construct an explicit model that produces such ``downliftings''. Once we fine tune the parameters such that the quadratic term in $\phi$ becomes small, the quartic term has to be taken into account in the analysis. This term is not independent from the mass term and the overall potential is phenomenologically interesting.

We conclude this appendix with a remark: the nice uplifting scenario overcomes the $\eta$ problem of KKLMMT allowing for fine tuning. Even if this is not the most satisfactory solution, it would be nevertheless interesting to study this kind of models to be able to explicitly quantify the required fine tuning. A naive, a priori (i.e. not based on any explicit setup) estimate of the required fine tuning might come out to be wrong in some explicit cases. An example is the hope to cancel the inflaton mass using the threshold corrections to the nonperturbative superpotential. Although this seemed a priori quite a reasonable expectation, direct investigation \cite{Krause:2007jk,Burgess:2006cb,Baumann:2007ah} has shown that, for two large classes of supersymmetric D7 embeddings, this is not possible\footnote{It might of course well be that other, yet not studied embeddings are suitable for the purpose.}.

%%%%%%%%%%%%%%%%%%%%%%%%%%%%%%%%%%%%%%%%%%%%%%%%%%%%%%%%%%%%%%%%%%%%%%%%%%%%%%%%%%%%%%%%%%%%%%%%%%%%%%%%%%%%

\section{From one to many}\label{app:many}

In this appendix we obtain the K\"ahler potential and the non perturbative superpotential when $N$ D3-branes are present.

\subsection{K\"ahler potential}

It was proposed in \cite{DeWolfe:2002nn} that in presence of a D3-brane, the K\"ahler potential should be modified to
\be\label{dewolfe}
\kq K=-3\mathrm{log}[\rho+\bar \rho-\gamma k(z,\bar z)],
\ee
where $\gamma$ a the constant describing the strength of gravitational backreaction, essentially proportional to $G_{N}$ (Newton constant) times the tension of the D3-brane $T_3$ and $k(z, \bar z)$ is the K\"ahler potential of the Calabi Yau evaluated at the position $z$ of the D3-brane. Two arguments were given to support this proposal. The first is based on the observation that the D3-brane in the presence of three form fluxes is BPS and does not feel any force. In addition, its inclusion should not spoil the no scale structure. This leads to the natural guess \eqref{dewolfe} with an arbitrary function $k(z,\bar z)$, for which one can easily check that the F-term potential vanishes, leaving $\rho$ and $z$ as flat directions.

The second argument consists in noticing that the kinetic term deduced from the supergravity approximation should reproduce, at leading order, the DBI action for the D3-brane. The 10-dimensional metric can be written as 
\be 
ds^2=h^{-1/2}(Y)e^{-6u}g_{\mu\nu}dx^{\mu}dx^{\nu}+h^{1/2}(Y)e^{2u}\tilde{g}_{ij}dY^idY^j,
\ee
so that we have singled out the breathing mode of the compact manifold $e^{2u}$ with respect to a fiducial metric $\tilde{g}_{ij}$. The expansion of the DBI action for small velocities gives
\be \label{DBI}
S_{\mathrm{DBI}}\simeq-T_3\int d^4x \sqrt{g}e^{-4u} \partial_{\mu} z^i\partial^{\mu} z^j \tilde{g}_{ij}.
\ee
To obtain this action from the second derivative of the K\"ahler potential as supergravity dictates, one is led to the identification of the breathing mode with the K\"ahler volume modulus plus the $X$ dependent shift \cite{DeWolfe:2002nn,Kachru:2003sx}:
\be 
e^{4u}=\rho+\bar \rho-\left(\frac{T_3}{3M_{Pl}^2}\right)k(z,\bar z).
\ee

To generalize to the case of $N_{D3}$ D3-branes it seems natural to use the same two criteria as above. A natural and simple guess is
\be\label{kmany}
\kq K=-3\mathrm{log}\left[ \rho+\bar \rho-\left(\frac{T_3}{3M_{Pl}^2}\right) \sum_{s=1}^{N_{D3}}  k(z_s,\bar z_s)\right],
\ee
where the index $s$ enumerates the D3-branes, located at $z_s$. This K\"ahler potential again respects the no scale structure. Also, after neglecting terms of order $e^{-8u}$, it gives a kinetic term that is the sum of $N_{D3}$ independent copies of the action \eqref{DBI}; this is what we expect, given that the D3-branes are mutually BPS. This property is important for the assisted inflation mechanism that requires the various inflatons to be weakly interacting.

%%%%%%%%%%%%%%%%%%%%%%%%%%%%%%%%%%%%%%%%%%%%%%%%%%%%%%%%%%%%%%%%%%%%%%%%%%%%%%%%%%%%%%%%%%%%%%%%%%%%%%%%%%%%%%%%%%%%%%%%%%%%%%%%%%%%%%%%%%%%%%%

\subsection{Superpotential}

In \cite{Baumann:2006th} the dependence of the non perturbative superpotential on the D3-brane position was calculated in the background of a warped throat. In the case of a Euclidean E3-brane or a stack of D7-branes with gaugino condensation, wrapping a divisor $\Sigma$ of the Calabi Yau, this superpotential is proportional to 
\be \label{Wnp}
W_{np}\propto e^{-T_3\,V^w_{\Sigma}/n},
\ee
where $V^w_{\Sigma}$ is the warped volume of the divisor $\Sigma$ (the label $w$ is to remind that the full metric, including the warp factor has to be use) ; $n$ is $1$ for E3-brane and $N_{D7}$ for gaugino condensation on a stack of $N_{D7}$ branes. As in the main text, we indicate with $g$ the holomorphic function whose zeros define the divisor $\Sigma$. 

It was argued in \cite{Ganor:1998ai} that $W_{np}$ has to depend on the D3-brane moduli. The argument consists in noticing that $W_{np}(z,\bar z)$ is the product of two factors, respectively a section of the line bundle and inverse line bundle defined by the divisor $\Sigma$ (wrapped by the D7's or the E3). Then a theorem guarantees that it has to vanish on (a surface homotopic to) the divisor $\Sigma$. 

The explicit dependence was first calculated in \cite{Berg:2004ek} for some toroidal orientifolds via a one loop open string calculation. In \cite{Baumann:2006th} the tree level closed string dual was considered. In the following we review the method of \cite{Baumann:2006th} and generalize the result to the case of $N_{D3}$ D3-branes. The idea is that a D3-brane backreacts on the metric, in particular changing the warp factor. This in turn changes the volume $V^w_{\Sigma}$ in \eqref{Wnp}. Schematically 
\be 
h(Y)&=&h_0(Y)+\delta h(Y,z)\, ,
\ee
where $Y$ is the point where the warping is evaluated, and $z$ is the D3-brane position. Integrating over the divisor $\Sigma$
\be  
V_{\Sigma}^w&=& (V_{\Sigma}^w)_0+\delta V_{\Sigma}^w(X)\\
&=&\int_{\Sigma}d^4Y \sqrt{g^{ind}(z,Y)}[h_0(Y)+\delta h(z;Y)]=\tilde{V_{\Sigma}}h_0(\rho+\bar \rho)+\Re \,\xi(z),\nonumber
\ee
where we have made explicit that the corrected volume is still the real part of a holomorphic function; $V^w_{\Sigma}$ is in fact proportional to the real part of the gauge kinetic function that supersymmetry dictates to be holomorphic. Finally
\be  
W_{np}&\propto& \mathrm{exp}\left(-\frac{T_3\xi(X)}{n}\right) \mathrm{exp} \left(-\frac{T_3 V_{\Sigma} h_0 \rho}{n}\right)\nonumber \\
&\equiv& A(X)e^{-a\rho}\,.
\ee
The advantage of this closed string approach is that $\delta h$ is directly calculated from the Laplace equation
\be 
-\nabla^2_Y \delta h (X,Y)= \kappa_{10} \rho_m
\ee
using the Green's function method. Here $\rho_m$ is the all inclusive energy density.

Solving the Laplace equation for $\delta h$ and integrating over a divisor defined by the zeros of the holomorphic function $g$, one obtains the dependence of the superpotential on the position $X$ of a single D3-branes \cite{Baumann:2006th}
\be 
W_{np}=A(z)e^{-a\rho}=A_0 g(z)^{1/n}e^{-a\rho},
\ee
where $A_0$ might depend on the dilaton and the complex structure moduli. To generalize to $N_{D3}$ D3-branes at positions $z_s$ one has to solve the Laplace equation
\be \label{laplacemany}
 - \nabla_Y^2 \delta h(z_s;Y) &=& 2T_3 \kappa_{10}^2
\left[\sum_{s=1}^{N_{D3}}\frac{\delta^{(6)}(z_s-Y)}{\sqrt{g(Y)}} - \rho_{bg}(Y) \right] \, \nonumber \\
  &=& 2T_3 \kappa_{10}^2
\left[\sum_{s=1}^{N_{D3}}\frac{\delta^{(6)}(z_s-Y)}{\sqrt{g(Y)}} - N\frac{\delta (z_0-Y)}{\sqrt{g(Y)}}\right] \, ,
\ee
where $z_0$ is a reference point (for example the tip of the cone). With this specific choice of $\rho_{bg}$ we get rid of a logarithmic divergence in $\delta \Vs$. This leads to an interpretation of $\delta h(X;Y)$ as the variation of the warp factor $h$ when the D3-branes are moved from $X_0$ to $X_s$.

Given the linearity of the Laplace equation the solution for \eqref{laplacemany} is just a sum over $X_s$ of the solution of the single D3-brane problem. Once $\delta h(X;Y)$ is integrated over the divisor $\Sigma$ and exponentiated, we obtain the superpotential 
\be  \label{supmany}
W_{np}=A(X)e^{-a\rho}=A_0 \prod_{s=1}^{N_{D3}} f(X_s)^{1/n}e^{-a\rho}\,.
\ee
We notice that the superpotential \eqref{supmany} reproduces what we would expect from Ganor's argument \cite{Ganor:1998ai}: $W_{np}$ vanishes if $\mathit{any}$ of the D3-branes hits the divisor $\Sigma$.

%%%%%%%%%%%%%%%%%%%%%%%%%%%%%%%%

\newpage
\section{Symbols Used in the Paper}
\label{sec:symbols}

\begin{table*}[!h]
%\caption{Symbols appearing in the paper}
\begin{center}
\begin{tabular}{ccll}
\hline
Var. & Mass dim. & Description & Definition\\
\hline \hline
$\Mp$& 1 & reduced Planck mass & $\Mp\equiv(8\pi G_{N})^{-1/2} $\\
$\kq$& -2 & & $\kq=\Mp^{-2}=8\pi G_N$\\
$z_A$ & -3/2 & complex conifold coordinates & $\sum_A z_A^2 = 0$\\
$x_A$ & -3/2 & real part of $z_i$ & \\
$y_A$ & -3/2 & imaginary part of $z_i$ & \\
$i$ &0& coordinate index & $i=1,2,3$\\
$A$ &0& coordinate index & $A=1,2,3,4$\\
$\zz$ &-3/2 & $\zz\equiv z_2^2+z_3^2$\\
$\xx$ &-3/2 & $\xx\equiv x_2^2+x_3^2$\\
$\varepsilon$ & -3/2 & conifold deformation parameter & $\sum_i z_i^2 = \varepsilon$ \\
$\mu$ & -3/2 & embedding parameter & $g(z_1)=\left(1 -z_1/\mu \right)$ \\
$r$ & -1 & radial coordinate on the conifold & $r^3 = \sum_i |z_i|^2$ \\
$K$ & 2 & moduli K\"ahler potential & $\kappa^2 {\cal K} = - 3 \log U$ \\
$k$ & -2 & conifold K\"ahler potential & \\
$W$ & 3 & superpotential & \eqref{sup} \\
$W_0$ & 3 & GVW-flux superpotential & $W_0 = \int G \wedge \Omega$\\
$W_{\rm np}$ & 3 & non-perturbative superpotential & $W_{\rm np} = A(z_i) e^{-a \rho}$\\
$A_0$ & 3 & prefactor of $W_{\rm np}$ & $A_0 = A(z_i =0)$ \\
$V_{F}$ &4& F-term potential & \\
$V(\phi)$& 4 & inflaton potential & \\
$\Lambda$&4 & term in the potential & \eqref{L} \\
$B$&4 & term in the potential & \eqref{B} \\
$C$&4 & term in the potential & \eqref{C} \\
$U$ & 0 & argument of K\"ahler potential & $U = \rho + \bar \rho - \gamma k$ \\
$\gamma$ & 2 & in sec. \ref{ST} and \ref{sec:kup}: factor in $K$   & $\gamma\equiv\sigma_0T_3/(3\Mp^2)$\\
$\gamma$ & 0 & in sec. \ref{sec:inlf} and \ref{no go}: inflaton Lorenz factor  & $\gamma\equiv1/\sqrt{1-f\dot{\phi}^2}$\\
$T_3$ & 4 & D3-brane tension &$(\alpha')^{-2}(2\pi)^{-3}g_s^{-1}$\\
$\rho$ & 0 & complex K\"ahler modulus & \\
$\sigma$ & 0 & real part of $\rho$ & $2 \sigma = \rho + \bar \rho$\\
$g(z)$ & & embedding equation & $A(z_i)\propto (g(z))^{1/n}$\\
$\g$ & & Kuperstein embedding's function & $g(z)=\g(\zz)-z_1$\\
$d$ & 1 & canonical normalization prefactor & $\varepsilon^{2/3}\sqrt{T_3c}$\\
$c$& 0 & numerical factor in $k$ & $2^{1/6}/3^{1/3}\simeq0.77$\\
$n$ & 0 & number of D7-branes & \\
$N_{D3}$ & 0 & number of D3-branes at the tip & \\
$N_e $ &0 & number of e-foldings &\\ 
\hline
\end{tabular}
\end{center}
\end{table*}
\begin{table*}
\begin{center}
\begin{tabular}{ccll}
\hline
Var. & Mass dim. & Description & Definition\\
\hline \hline
$a$ & 0 & parameter in $W_{np}$ & $a\equiv2\pi/n$ \\
$f$ & -4 & appears in the DBI-like kinetic term &  \\
$\eta$& 0 & slow roll parameter & $\Mp^{-2}V''/V$\\
$\epsilon$& 0 & slow roll parameter & $\Mp^{-2}(V'/V)^2$\\
\hline
\end{tabular}
\end{center}
\end{table*}

\end{document}